\documentclass[
reprint,
superscriptaddress,
amsmath,
amssymb,
aps,
pra
]{revtex4-2}
\usepackage{tabularx}
\usepackage{adjustbox}
\usepackage{longtable}
\usepackage{makecell}  
\usepackage{graphicx}
\usepackage{dcolumn}
\usepackage{bm}
\usepackage[colorlinks, linkcolor=blue, anchorcolor=blue, citecolor=blue]{hyperref}
\usepackage{color}
\usepackage{xcolor}

\begin{document}

\preprint{APS/123-QED}

\title{Calibrating the Role of Entanglement in Variational Quantum Algorithms from a Geometric Perspective}


\author{Chunxiao Du}
\affiliation{School of Physics, Beihang University, Beijing 100191, China}

\author{Yang Zhou}
\email[Corresponding author: ]{yangzhou9103@buaa.edu.cn}
\affiliation{School of Physics, Beihang University, Beijing 100191, China}

\author{Zhichen Huang}
\affiliation{School of Physics, Beihang University, Beijing 100191, China}

\author{Rui Li}
\affiliation{School of Applied Science, Beijing Information Science and Technology University. Beijing 100192 China}

\author{Zheng Qin}
\affiliation{Shenzhen Institute of Beihang University. Shenzhen 518063 China}

\author{Shikun Zhang}
\affiliation{School of Future Technology, Henan University, Kaifeng, China}

\author{Zhisong Xiao}
\affiliation{School of Physics, Beihang University, Beijing 100191, China}
\affiliation{School of Instrument Science and Opto-Electronics Engineering, Beijing Information Science and Technology University, Beijing 100192, China}

\date{\today}

\begin{abstract}
Calibrating the role of entanglement in quantum algorithms is a crucial task in the development of quantum computing. Most existing studies have primarily focused on how the static properties of entanglement—such as its magnitude and phase—affect key performance metrics. In this work, we instead explore the relationship between the dynamical behaviors of entanglement and the execution of variational quantum algorithms from a geometric perspective.  We find that, in contrast to conventional Hamiltonian dynamics where the evolution process is dominated by the dynamical phase, quantum state evolution in quantum algorithms is primarily governed by the geometric phase with the trajectory determined by the parameter-dependent Hilbert space geometry. In the problem-agnostic Hardware-Efficient Ansatz (HEA), entanglement dynamics and state evolution are decoupled. Conversely, in the problem-inspired Hamiltonian Variational Ansatz (HVA), the dynamical phase contribution is enhanced, allowing entanglement to function as a dynamical resource: more entanglement consumption correlates directly with faster quantum state evolution.
\end{abstract}

\maketitle


\section{\label{sec:level1}Introduction}

Quantum computing~\cite{1982Simulating} is widely regarded as capable of solving practical, real-world problems more effectively than classical computing, which is referred to as ``quantum advantage"~\cite{2019Instead}. A fundamental yet unresolved question lies in how such an advantage arises from the distinctive properties of quantum formalism, such as quantum correlation~\cite{1935Can}, non-stabilizerness~\cite{2005Universal}, and contextuality~\cite{1967The}. Among these, quantum entanglement~\cite{Schrodinger1935EPR}, one of the most prominent features of quantum mechanics, is considered a core resource for quantum computing and is well recognized as essential for achieving quantum advantage. However, despite its recognized significance, the precise role entanglement plays in the functioning of quantum algorithms remains an open topic of debate.

Early studies generally suggested that "more entanglement" implies "greater computational power." Specifically, if a quantum state contains no entanglement, the corresponding algorithm can be simulated by classical methods; only by introducing a sufficient amount of entanglement can quantum advantage be achieved~\cite{Hayden_2004, hayden2006aspects}. However, further research revealed a more nuanced picture. In certain quantum algorithms, states with only mild entanglement can still support universal quantum computation without significantly compromising efficiency~\cite{gross2009most, 2013Universal, 2023Quantum}. Conversely, excessive entanglement may hinder the realization of quantum speedup. For instance, recent studies have uncovered a dual role of entanglement in variational quantum algorithms (VQAs): while moderate entanglement enhances algorithmic expressivity, excessive entanglement can cause the gradient of the cost function to decay exponentially with system size, leading to the emergence of barren plateaus (BP)~\cite{wiersema2020exploring, PRXQuantum.2.040316, 2020Entanglement, zhang2024absence, zhang2024single, 2025Machine}. Moreover, it has been revealed that the entanglement entropy of most quantum algorithms complies with a volume law, posing a major obstacle to classical simulability~\cite{Dupont_2022, Dupont_2022Calibrating, nakhl2024calibrating, 2017Quantum}.

Most existing research has predominantly focused on how static properties of entanglement—such as its magnitude and phase—in the limit of large circuit depth affect key performance metrics of quantum algorithms~\cite{Cerezo_2021, zhao2025entanglement, wiersema2020exploring, D_ez_Valle_2021}, including expressibility, trainability, and classical simulability. However, little attention has been paid to the role of the dynamic characteristics of entanglement, particularly how its growth and spread throughout real algorithm execution influence algorithmic performance. In this work, we investigate this question within the framework of VQAs~\cite{peruzzo2014variational}, which are widely regarded as the most promising candidates for achieving near-term quantum advantage. For contrast, two different circuit structures, hardware-efficient ansätze (HEA)~\cite{kandala2017hardware} and Hamiltonian variational ansätze (HVA)~\cite{wecker2015progress}, are employed to solve the ground state energy of the Ising model. The evolution of entanglement entropy is traced by its layer-by-layer variation, and the execution of the quantum algorithm is characterized through the geometry of quantum state evolution.

Under the geometric framework, we monitor the evolution of two complementary geometric quantities: the geodesic distance towards the target space and the geodesic distance from the initial state. The former characterizes the effective proximity to the desired solution subspace, thus serving as a direct metric of computational progress and convergence efficiency. The latter has been established to govern the fractional contribution of the dynamical and geometric phases to the total phase, while simultaneously capturing the excess evolution of the quantum state through the projective Hilbert space relative to the optimal geodesic trajectory~\cite{pati2025fractionalcontributiondynamicalgeometric}. By correlating the step-wise variation in entanglement with changes in both geodesic distance towards the target space and geometric phase fraction, we are able to discern whether dynamical variation in entanglement actively drives state evolution or merely accompanies it as a passive byproduct. Both initial randomized circuits and final optimized circuits are considered. Our numerical simulations reveal two distinct pictures: in HEA circuits, the evolution of the quantum state is overwhelmingly dominated by the geometric phase, and entanglement dynamics are consistently decoupled from state evolution across both random and optimized regimes; conversely, in HVA circuits, the contribution from the dynamical phase is enhanced and a robust positive correlation emerges—greater entanglement consumption directly translates into accelerated quantum state evolution. These findings demonstrate that the capacity to harness entanglement as a dynamical resource is not inherent to all quantum circuits, but is instead contingent on the problem-informed inductive bias of the ansatz architecture.

The rest of the article is organized as follows. In Sec.~\ref{sec:levelP}, we introduce the definitions and methods used throughout this work. In Sec.~\ref{sec:levelR}, we present our main results, analyzing the entanglement dynamics and geodesic distance behavior in both the initial randomized circuits and the final optimized circuits. Finally, in Sec.~\ref {sec:levelC}, we summarize our findings and discuss their implications. 

\section{\label{sec:levelP}PRELIMINARY}

\subsection{\label{sec:levelEM}Entanglement Entropy}

We use the bipartite Von Neumann entanglement entropy to quantify the quantum entanglement generated during the execution of the quantum algorithm. For a parameterized quantum circuit $U\left(\boldsymbol{\theta}\right)$ acting on an $N$-qubit Hilbert space, the corresponding quantum state is represented as
\begin{equation}
    |\psi(\boldsymbol{\theta})\rangle = U(\boldsymbol{\theta}) |0\rangle^{\otimes n}.
    \label{eq1}
\end{equation}
Given a bipartition of the system into a subsystem $A$ and its complement $B$, the reduced density matrix of subsystem $A$ is obtained by tracing out the degrees of freedom associated with $B$
\begin{equation}
    \rho_A(\boldsymbol{\theta}) = \mathrm{Tr}_B \Bigl[\, |\psi(\boldsymbol{\theta})\rangle \langle \psi(\boldsymbol{\theta})| \,\Bigr].
    \label{eq2}
\end{equation}
The Von Neumann entanglement entropy is then defined as
\begin{equation}
    S_A(\boldsymbol{\theta}) = -\mathrm{Tr}\Bigl[\, \rho_A(\boldsymbol{\theta}) \log \rho_A(\boldsymbol{\theta}) \,\Bigr] = - \sum_i \lambda_i(\boldsymbol{\theta}) \log \lambda_i(\boldsymbol{\theta}),
    \label{eq3}
\end{equation}
where $\{\lambda_i(\theta)\}$ are the eigenvalues of $\rho_A(\theta)$.

Throughout our analysis, the maximal bipartite entanglement entropy is used as a quantitative measure of entanglement produced by the parameterized circuits. Balanced bipartitions, including the half-half case for even-qubit systems, are known to yield the greatest entanglement among all possible divisions. Thereby, for an $N$-qubit state, we choose subsystem sizes $|A| = \left\lfloor \frac{N}{2} \right\rfloor$(the largest integer smaller than $\frac{N}{2}$) and $|B| = \left\lceil \frac{N}{2} \right\rceil$(the smallest integer greater than $\frac{N}{2}$). To simulate the quantum state evolution and compute the entanglement entropy efficiently, the matrix product state (MPS)~\cite{vidal2004efficient, schollwock2011density, orus2014practical} representation is employed. It allows us to represent quantum states with limited entanglement using a polynomial number of parameters, enabling efficient computation of reduced density matrices and entanglement measures even for relatively large systems.




\subsection{\label{sec:levelGD}Geometric Perspective}

We know that the execution of every quantum algorithm can be interpreted as a quantum state evolution that drives a specific initial state $|\psi_0\rangle$ to a target state $|\psi_T\rangle$ that encodes a problem solution. Geometrically speaking, such a state evolution traces a path in the projective Hilbert space. The total distance travelled by the quantum state can be measured using the Fubini-Study distance~\cite{1981Statistical}, which quantifies the instantaneous separation between quantum states in the projective Hilbert space. The accumulation of the Fubini-Study distance $S(|\psi_0\rangle, |\psi_T\rangle )$ is the actual distance travelled. This geometric perspective reveals that quantum state evolution can be viewed as a trajectory through curved state space. Thus, there should exist a shortest path distance between the initial and the target states, while the actual evolution typically follows a longer path. In geometrical terms, the shortest, most direct path of this state evolution would be characterised by the geodesic from the initial state to the target. The geodesic distance between two normalized quantum states is given by~\cite{anandan1990geometry}
\begin{equation}
    S_0\bigl( |\psi_1\rangle, |\psi_2\rangle \bigr) = 2\arccos \Bigl( \bigl| \langle \psi_1 | \psi_2 \rangle \bigr| \Bigr).
    \label{eq4}
\end{equation}

Denote $|\psi_t\rangle$ as an intermediate state during the evolution from the initial state $|\psi_0\rangle$ to the target state $|\psi_T\rangle$. According to Ref.~\cite{pati2025fractionalcontributiondynamicalgeometric}, the geodesic distance $S_0\bigl( |\psi_0\rangle, |\psi_t\rangle \bigr)$ has twofold physical implications. Firstly, it governs the fractional contribution of the dynamical phase $\Phi_D$ and the geometric phase $\Phi_G$ to the total phase $\Phi_T$. The geometric phase fraction is defined as
\begin{equation}
    f_g(t)\equiv{f_g(|\psi_t\rangle)}={{\textit{d}\Phi_G}\over{\textit{d}\Phi_T}}=\sin^2\bigl({{S_0\bigl( |\psi_0\rangle, |\psi_t\rangle \bigr)}\over2}\bigr).
    \label{fg}
\end{equation}
It can be seen that when $S_0$ is small, little geometric phase accumulates per unit total phase, while when $S_0$ approaches $\pi$, almost all total phase change converts into geometric phase, reflecting the dominant role of state-space curvature in dynamics. Secondly, this factor signifies how much the geodesic path at each instant differs from the actual path that the system travels, since the infinitesimal Fubini-Study distance can be given by
\begin{equation}
 \textit{d}S^2=\textit{d}S_0^2+\sin{S_0^2}\textit{d}\Phi_T^2 .
    \label{fs}
\end{equation} 
The quantity $\textit{d}S-\textit{d}S_0$ captures the extra evolution of the quantum state through the projective Hilbert space.     

Apart from the geodesic distance from the initial state $|\psi_0\rangle$ to an intermediate state $|\psi_t\rangle$, we also compute the geodesic distance from an intermediate state $|\psi_t\rangle$ to the target state $|\psi_T\rangle$. This additional measure helps characterize the efficiency of the corresponding quantum algorithm execution. It should be noted that for a computational problem, there are usually more than one solution state. A quantum algorithm has the freedom to reach any state within the target solution space, which contains all superpositions of quantum states encoding problem solutions. Accordingly, the geodesic distance from an intermediate state $|\psi_t\rangle$ to the target solution space $\mathcal{T}$ can be defined as the shortest geodesic distance to one of the states in the target solution space. The solution space eigenbasis is determined by solving the target Hamiltonian $H$ via exact diagonalization as
\begin{equation}
    H |v_k\rangle = E_0 |v_k\rangle, \quad k = 1,2,\dots,m,
    \label{eq5}
\end{equation}
where ${E_0}$ are the ground state energy and $|v_k\rangle$ form an orthonormal basis of the solution space $\mathcal{T}$. Then, any target solution state can be written as
 \begin{equation}
    |\psi_T\rangle = \sum_{k=1}^{m} c_k |v_k\rangle,
    \label{eqT}
\end{equation}
where $c_k$ is the complex superposition coefficient. 
Given a quantum state $|\psi_t\rangle$, its proximity to the space $\mathcal{T}$ is quantified by evaluating its maximal overlap with the target solution states
\begin{equation}
    s(|\psi_t\rangle, \mathcal{T}) = \max_{\{\mathcal{T}\}} \bigl| \langle \psi_t | \psi_T \rangle \bigr|.
    \label{eq7}
\end{equation}
Define the projection operator onto $\mathcal{T}$ as
\begin{equation}
   \boldsymbol{P} = \sum_{k=1}^{m} |v_k\rangle\langle v_k |.
    \label{eq6}
\end{equation}
Eq.~(\ref{eq7}) is equivalent to~\cite{kruger2025quantum}
\begin{equation}
    s(|\psi_t\rangle, \mathcal{T}) = \sqrt{\langle \psi_t | \boldsymbol{P} |\psi_t\rangle}.
    \label{eq8}
\end{equation}
This overlap $s \in [0, 1]$ captures the degree to which $|\psi\rangle$ resides in the solution space and generalizes the state-to-state geodesic distance to a state-to-space formulation.


\subsection{\label{sec:levelTFIM}Transverse-field Ising Model}

In this work, we consider the one-dimensional transverse-field Ising model (TFIM), a paradigmatic quantum many-body model that captures the competition between interaction-induced order and field-induced quantum fluctuations. The Hamiltonian is given by
\begin{equation}
    H = -J \sum_{j=1}^{N} \sigma_j^z \sigma_{j+1}^z - h \sum_{j=1}^{N} \sigma_j^x,
    \label{eq9}
\end{equation}
where $\sigma_j^\alpha \ (\alpha = x, y, z)$ denotes a Pauli operator acting on site $j$. A periodic boundary condition with $\sigma_{N+1}^z\equiv\sigma_1^z$ is employed throughout this work unless otherwise specified.

The first term represents nearest-neighbor Ising interactions with coupling strength $J$, which favors ferromagnetic ordering along the $z$-direction. The second term introduces a transverse magnetic field of strength $h$, which drives spin flips along the $x$-direction and induces quantum fluctuations. The interplay between these two competing terms results in a quantum phase transition in the critical field $h_c = J$, separating a ferromagnetic phase $(h<h_c)$ from a paramagnetic phase $(h> h_c)$. In this work, we set $h=J=1$. The corresponding system is gapless in the thermodynamic limit. Its ground state is a highly entangled critical state with long-range correlation.


\subsection{\label{sec:Qc}Quantum Circuit Structures}

To obtain the ground-state energy of the transverse-field Ising Hamiltonian, we employ the variational quantum eigensolvers (VQEs), which use parametrized quantum circuits (PQCs) to run on a quantum computer and then outsource the parameter optimization to a classical optimizer. In this work, two distinct circuit structures are explored.

\begin{figure}[h]
    \centering
    \includegraphics[width=0.45\textwidth]{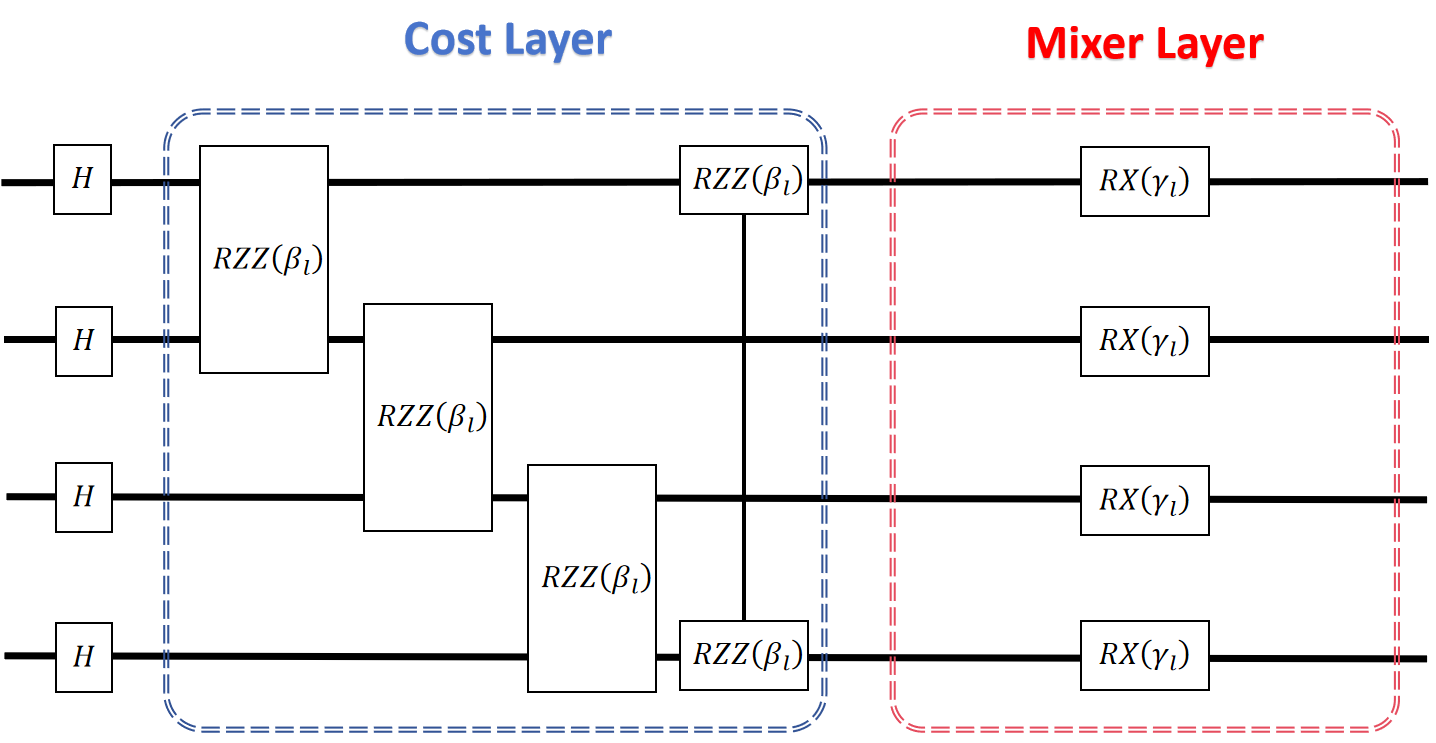}
    \caption{The HVA quantum circuit structure for the TFIM with $N=4$ and $L=1$.}
    \label{fig1}
\end{figure}

The first circuit structure is the HVA, constructed by exponentiating terms that mirror the structure of the target Hamiltonian. For an $N$-qubit system, a single layer of the HVA circuit takes the form of
\begin{equation}
    U_{\rm HVA}(\beta_l, \gamma_l ) = \prod_{j=1}^{N} \exp\Bigl( -i \frac{\beta_l}{2} Z_j Z_{j+1} \Bigr)\prod_{j=1}^{N} \exp\Bigl( -i \frac{\gamma_l}{2} X_j \Bigr),
    \label{eq10}
\end{equation}
where $l$ represents the $l$-th layer, and $j$ represents the $j$-th qubit. $\beta_l$ and $\gamma_l$ are layer-dependent variational parameters associated with the interaction and transverse-field terms, respectively. For a circuit with $L$ layers, the full unitary matrix is
\begin{equation}
    U_{\rm HVA}(\boldsymbol{\beta}, \boldsymbol{\gamma}) = \prod_{l=1}^{L} U_{\rm HVA}(\beta_l, \gamma_l)
    \label{eq11}
\end{equation}
yielding a total of $2L$ variational parameters. Fig.~\ref{fig1} illustrates the corresponding quantum circuit for $N=4$ and $L=1$.
The leftmost Hadamard gates, represented by $H$, are used to construct the initial $|+\rangle$ state. The middle blue-shaded region represents the local Ising-type interactions implemented by two-qubit rotation gates $R_{ZZ}(\beta_l) = \exp\!\left(i\frac{\beta_l}{2} \sigma_j^z \sigma_{j+1}^z \right)$. Such a layer is generally termed the cost layer as it represents the unitary evolution generated by a component of the cost Hamiltonian. The right red-shaded region represents the transverse field coupling realized via single-qubit rotation gates $R_X(\gamma_l) = \exp\!\left(i\frac{\gamma_l}{2} \sigma_j^x \right)$. Such a layer is generally termed the mixer layer as its role is to redistribute probability amplitudes across different computational basis states. 

\begin{figure}[h]
    \centering
    \includegraphics[width=0.45\textwidth]{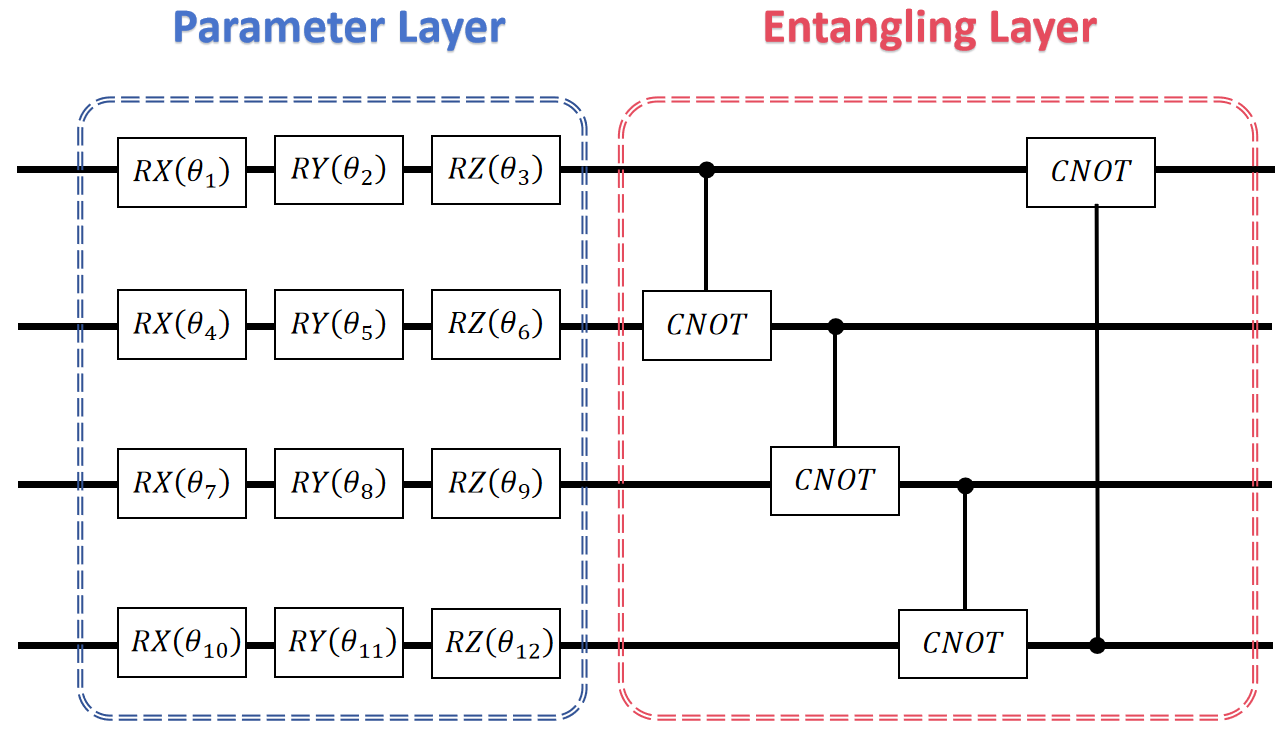}
    \caption{The HEA quantum circuit structure with $N=4$ and $L=1$.}
    \label{fig2}
\end{figure}

The second circuit structure is the HEA, which is constructed using gates that can be directly executed on near-term quantum hardware. A typical HEA consists of single-qubit gate layers with tunable parameters and two-qubit gate layers to provide entanglement. This can be represented as follows
\begin{eqnarray}
  U_{\rm HEA}(\boldsymbol{\theta}) =\prod_{l=1}^{L}U(\theta_{l})W_{l}
  \label{equ1}
\end{eqnarray}
with 
\begin{eqnarray}
  U(\theta_{l}) =\prod_{j=1}^{N} \exp\Bigl( -i\theta_{l}^{j}\sigma/2\Bigr)              
  \label{equ2}
\end{eqnarray}
where $\sigma \in (\sigma_{x},\sigma_{y},\sigma_{z})$ is one of the Pauli matrices. $W_{l}$ are unparametrized two-qubit gates. Fig.~\ref{fig2} illustrates the corresponding circuit for $N=4$ and $L=1$. The blue-shaded block on the left represents the rotation layer in which each qubit undergoes three independent single-qubit rotations around the $X$, $Y$, and $Z$ axes, providing a flexible local parameterization consisting of $3n$ rotation parameters per layer. The right red-shaded block represents an entangling layer implemented by a brickwall linear arrangement of $CNOT$ gates. For a circuit with $L$ layers, the total number of parameters is  $3nL$.  


Although both ansätze can be used to obtain the ground-state energy of quantum many-body systems, they differ significantly in the circuit design. The HVA structure is model-specific and incorporates the Hamiltonian terms with a compact two-parameter-per-layer design. In contrast, the HEA is model-agnostic and generates a larger Hilbert space, exhibiting a more complex optimization landscape. These structural differences lead to distinct behaviors in convergence, entanglement generation, ground-state energy accuracy, and quantum state evolution.

\section{\label{sec:levelR}MAIN RESULTS}

All results presented in this section are obtained through numerical simulations under the MPS framework. We consider two regimes: a relatively large-depth regime, with $N=10$ qubits and $L=20$ layers for HVA and $N=6$ qubits and $L=20$ layers for HEA, which can approximate the behaviors in the thermodynamic limit; and a practical regime, with $N=10$ qubits and $L=10$ layers for HVA and $N=6$ qubits and $L=9$ layers for HEA, which can obtain the optimal ground-state energy. For each configuration, 1000 independent trials with random parameter initializations drawn from the Haar distribution are performed. The optimization is carried out for 200 iterations, targeting the minimization of the expectation value of the TFIM Hamiltonian with parameters $J=1$ and $h=1$. We analyze both the initial randomized circuits and the final optimized circuits to explore entanglement dynamics and the evolution of quantum states during algorithm execution.

We simulate our circuits using the PennyLane package and a Python environment. The classical optimizer is Adam~\cite {2014Adam} with a learning rate of 0.5.

\subsection{\label{sec:levelINIT}Results of Initial Randomized Circuits}

For any VQE algorithm, a standard starting point is to initialize the circuits with random parameters sampled from the Haar distribution. Therefore, we first investigate the role of entanglement in both HVA and HEA executions with initial randomized circuits.



\begin{figure}[h]
    \centering

    \includegraphics[width=0.35\textwidth]{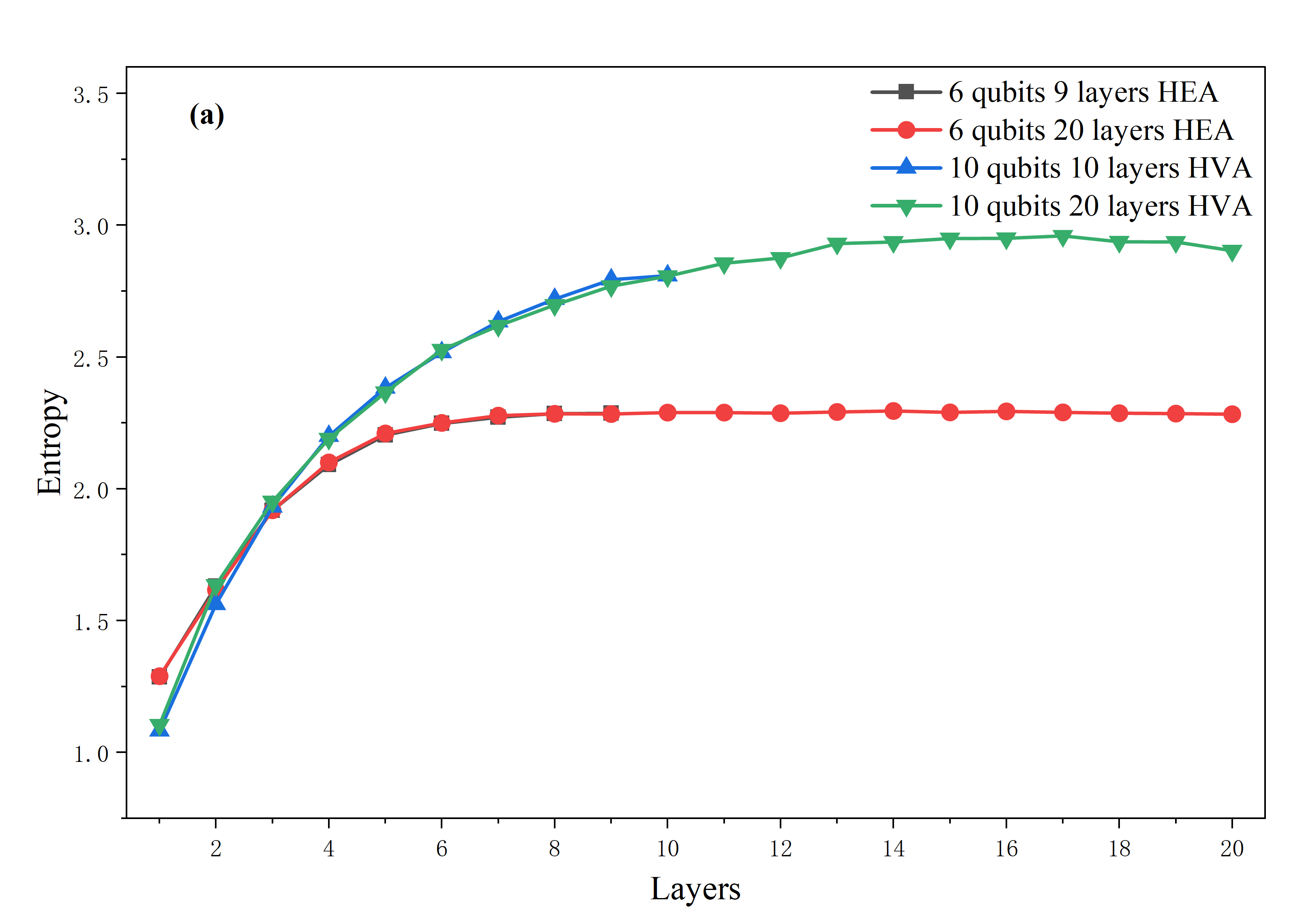}
    \includegraphics[width=0.35\textwidth]{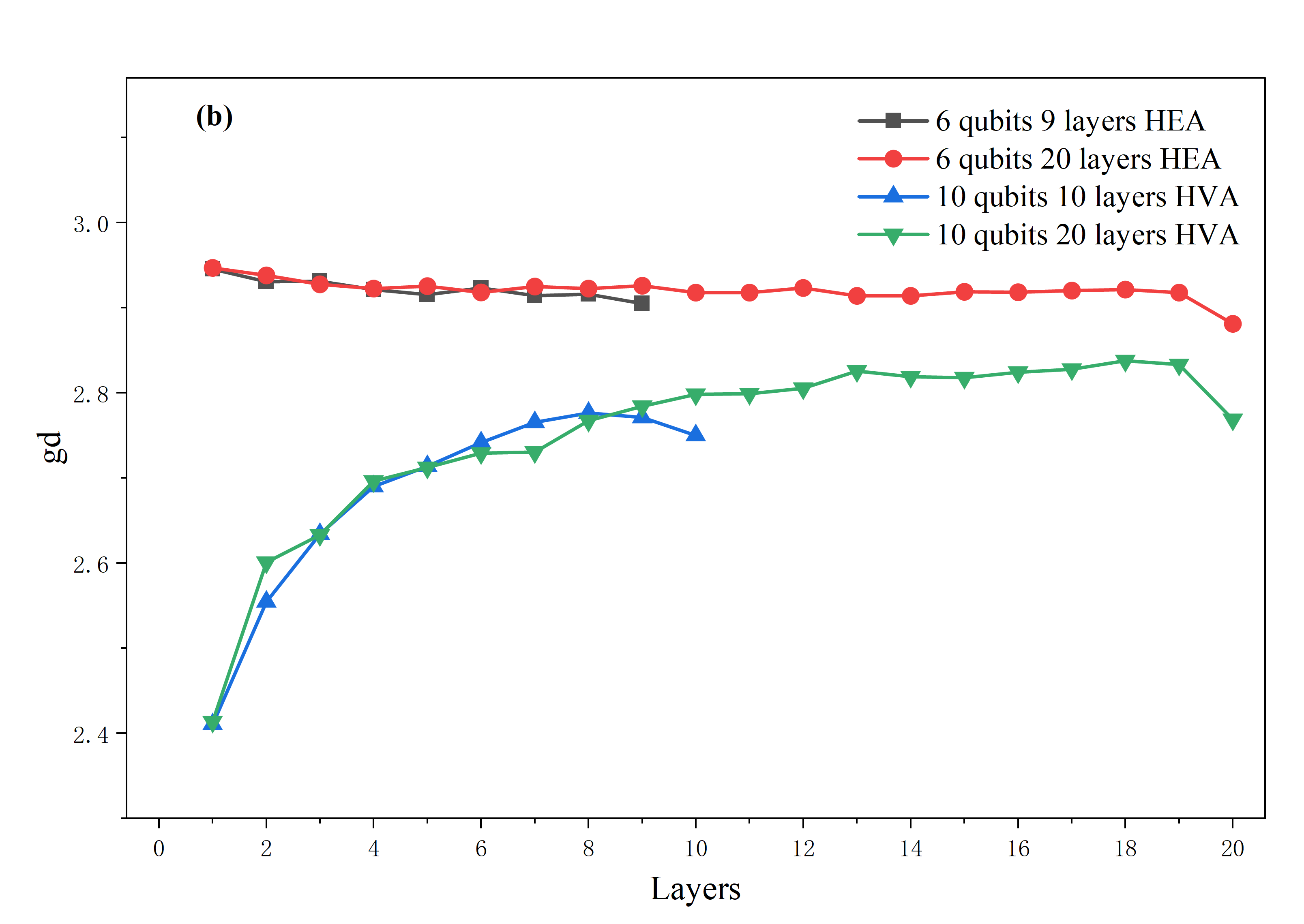}
    \includegraphics[width=0.35\textwidth]{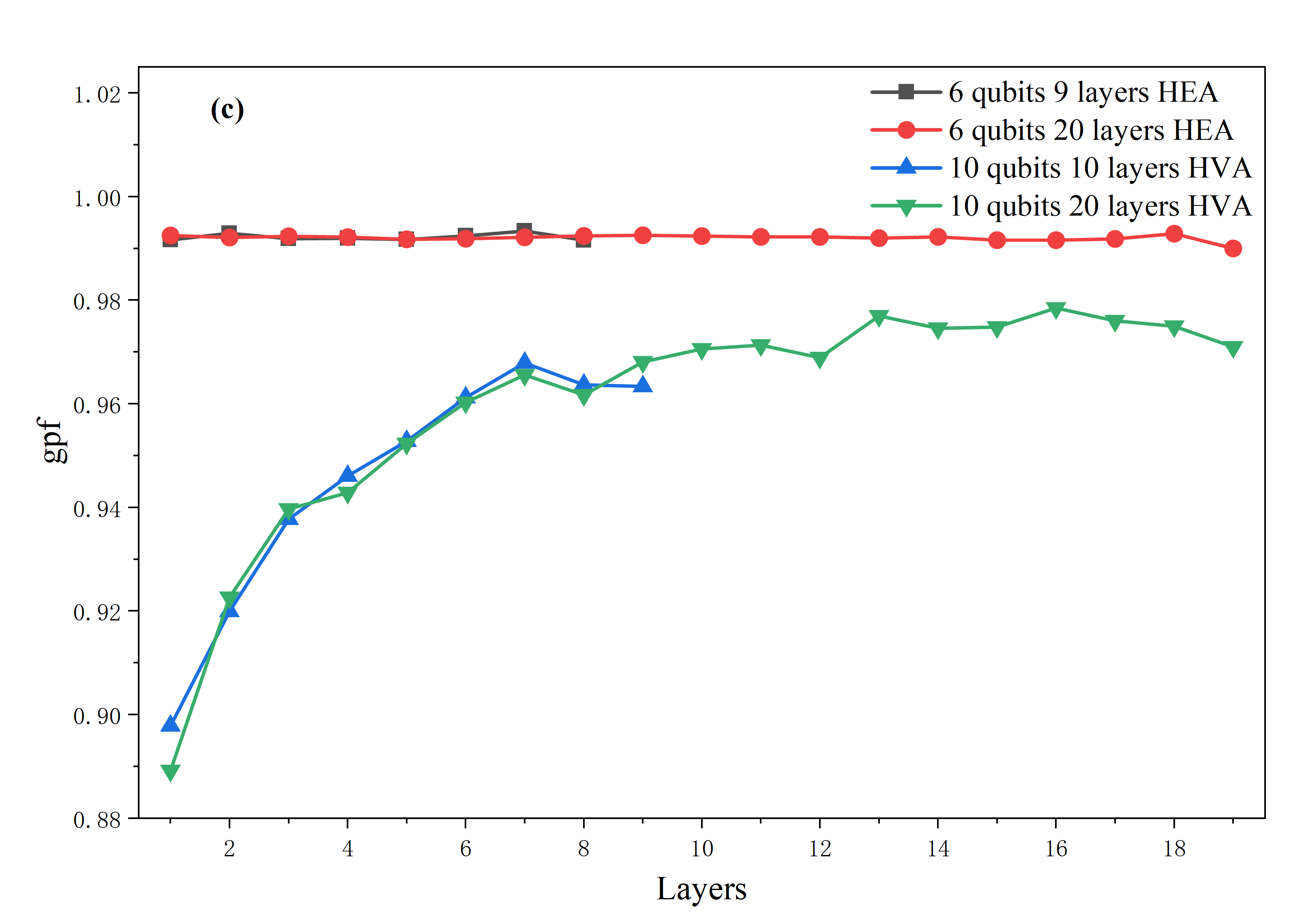}

    \caption{Data averaged $10^3$ independent trials with random parameter initializations for both HVA and HEA circuits. (a) Variations of the bipartite entanglement entropy as a function of circuit layer. (b) Variation of geodesic distance between intermediate states and the target solution space as a function of circuit layer. (c) Variation of geometric phase fraction as a function of circuit layer.}
    \label{fig3}
\end{figure}

Fig.~\ref{fig3}(a) reveals that, for both HVA and HEA, the bipartite entanglement entropy increases monotonically with the number of layers and eventually saturates at a high value under both conditions. The corresponding variation of geodesic distance from intermediate states towards the target space is shown in Fig.~\ref{fig3}(b), which reflects the properties of quantum states evolution during the execution of algorithms. It can be observed that for HEA, the geodesic distance $gd$ to the target space remains consistently large across all layers. In contrast, for HVA, this distance begins at a relatively close value and then progressively increases as the number of layers grows. Besides, the HVA circuit yields a slightly smaller geodesic distance than the HEA at equivalent depths, even when implemented at a larger scale.

To further investigate the physical factors governing the evolution of the quantum state during different algorithm executions, we also computed the geodesic distance between the intermediate states and the initial state. Based on Eq.~(\ref{fg}), the geometric phase fraction $gdf$ was calculated for each layer, and the results are presented in Fig.~\ref{fig3}(c). It is evident that for the HEA circuits, the geometric phase fraction remains close to unity across all layers. This indicates that the state evolution is predominantly driven by the geometric configuration of the Hilbert space rather than by dynamical contributions. In stark contrast, the HVA circuit exhibits a markedly lower geometric phase fraction in the initial layers, suggesting that its early evolution is significantly influenced by dynamical factors aligned with the target Hamiltonian, before gradually transitioning toward a regime where geometric effects become more pronounced at greater depths. This trend indicates that with the progressive accumulation of randomness introduced by successive parameterized gates, the influence of the problem-inspired inductive bias is effectively diluted. Consequently, the HVA circuit loses its initial dynamical guidance and begins to explore the Hilbert space in a manner more akin to the unstructured HEA, where the trajectory of the quantum state is governed by the geometric curvature of the Hilbert space.



\begin{figure}[h]
    \centering
    \includegraphics[width=0.235\textwidth]{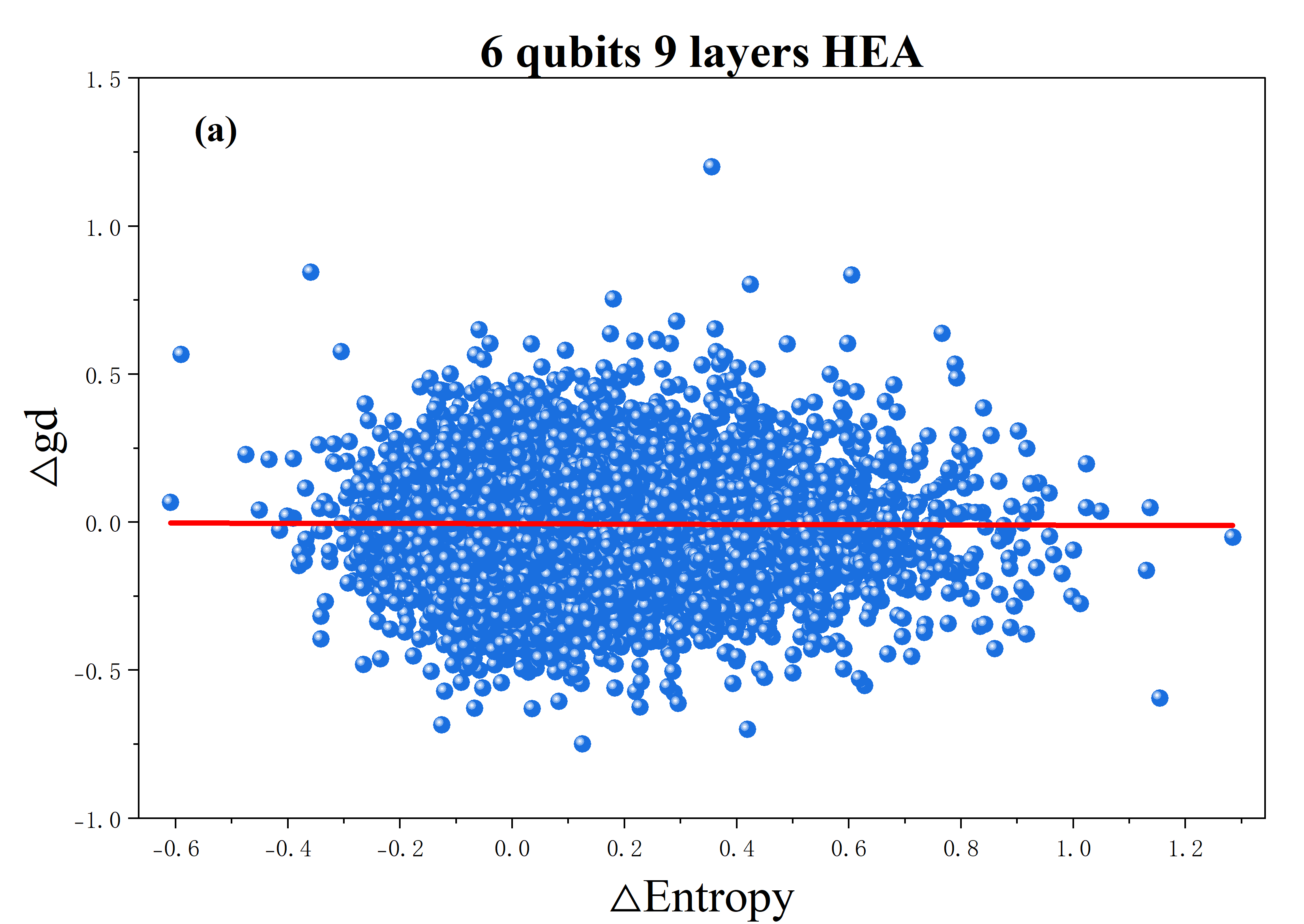}
    \includegraphics[width=0.235\textwidth]{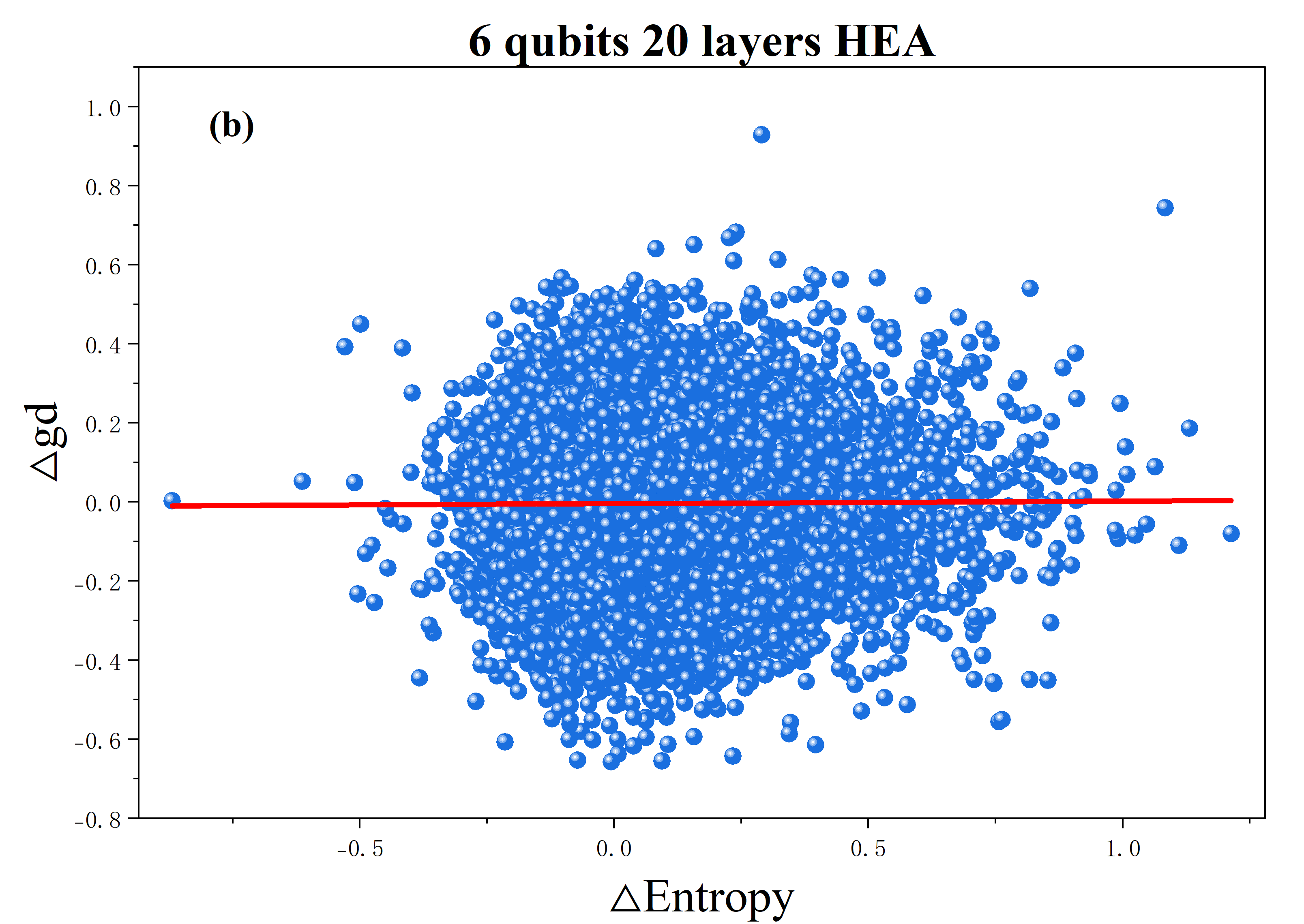}

    \vspace{0.25cm}

    \includegraphics[width=0.235\textwidth]{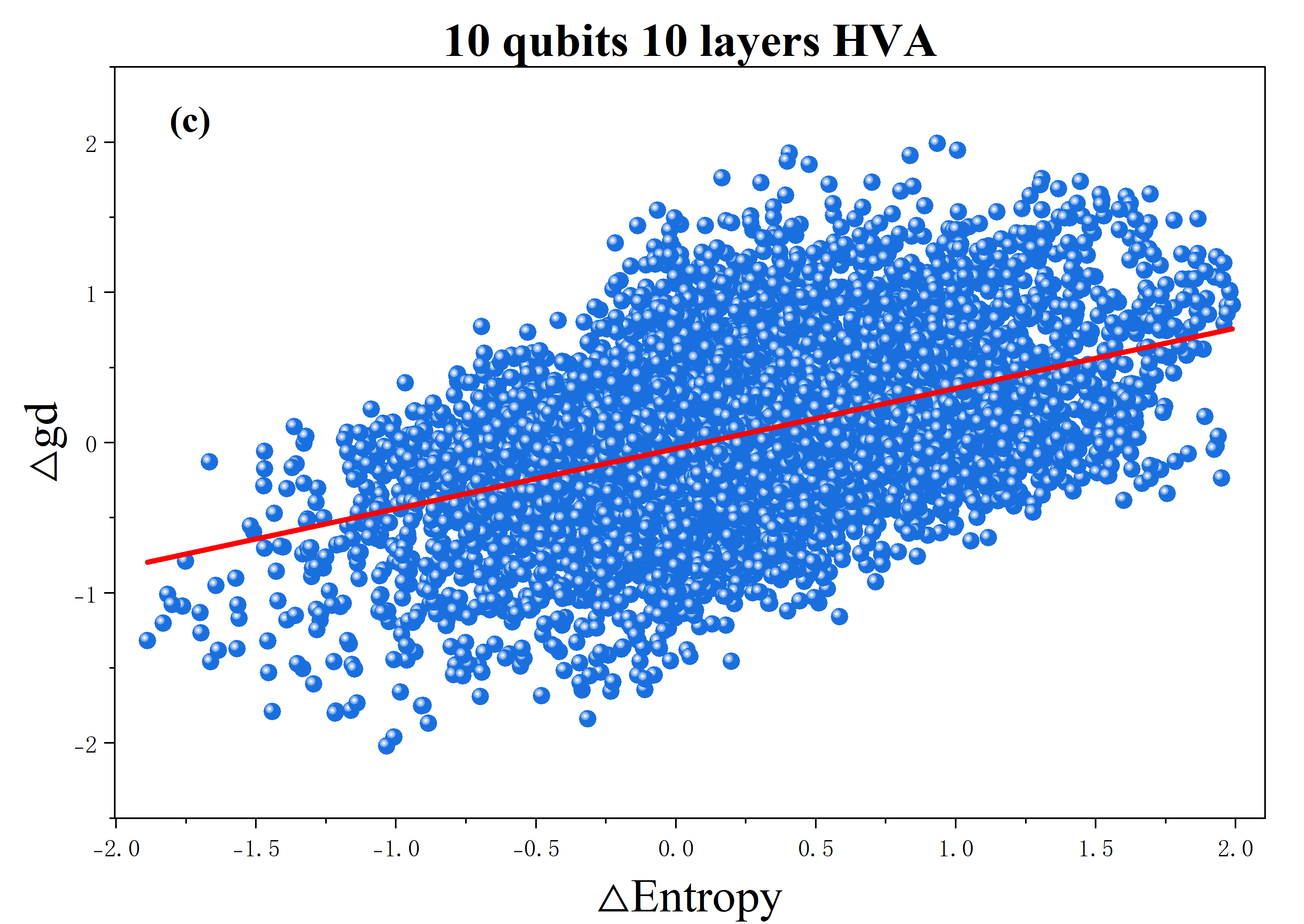}
    \includegraphics[width=0.235\textwidth]{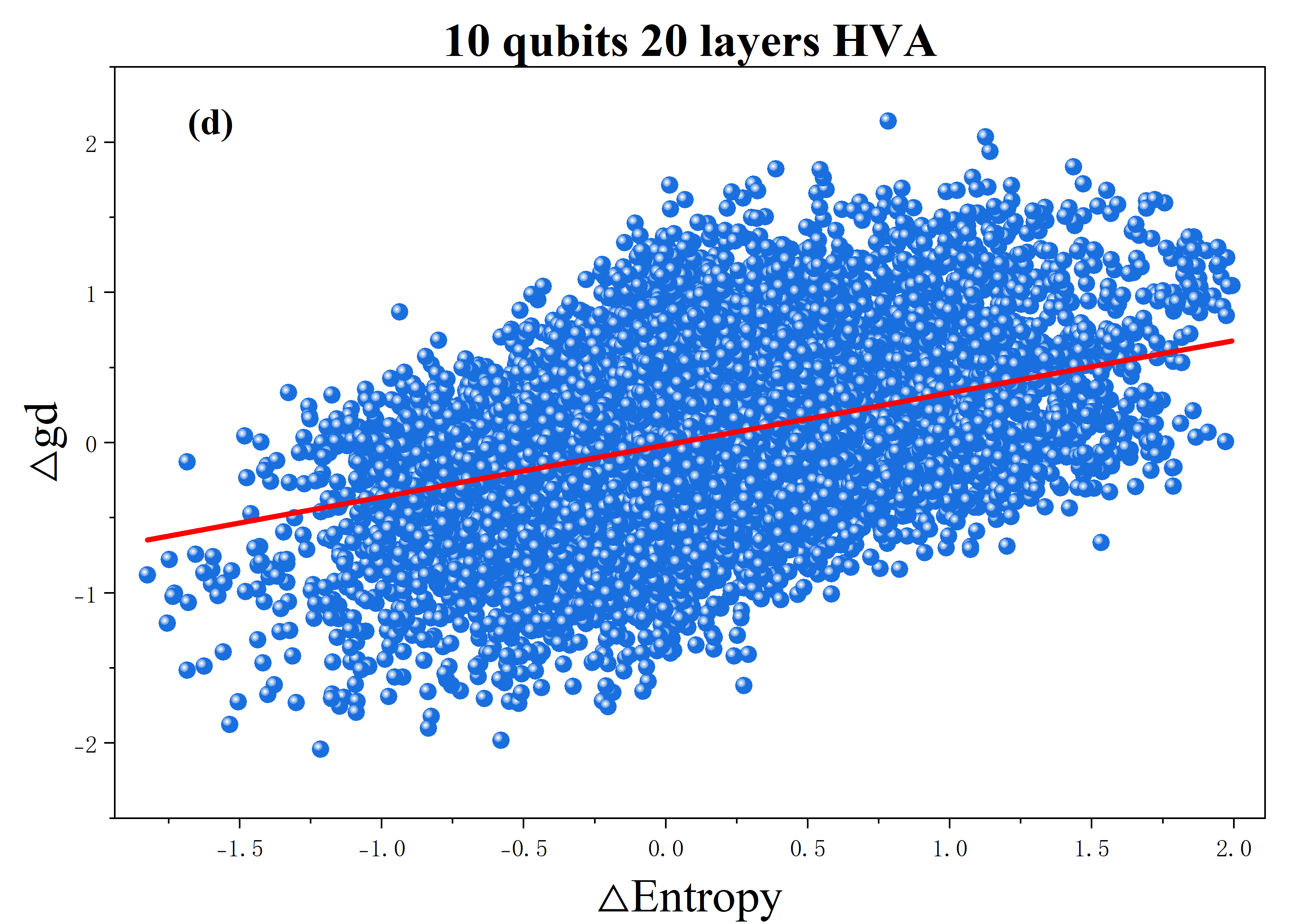}

    \caption{Statistical correlations between changes of entanglement entropy and geodesic distance $gd$ toward the target space. The variation is calculated between adjacent layers across $10^3$ randomized HVA circuits and $10^3$ randomized HEA circuits. The red lines represent the relationship between layer-by-layer changes of the two quantities. (a) and (b) show the results of HEA circuits, while (c) and (d) show the results of HVA circuits.}
    \label{fig4}
\end{figure}

\begin{figure}[htbp]
    \centering
    \includegraphics[width=0.235\textwidth]{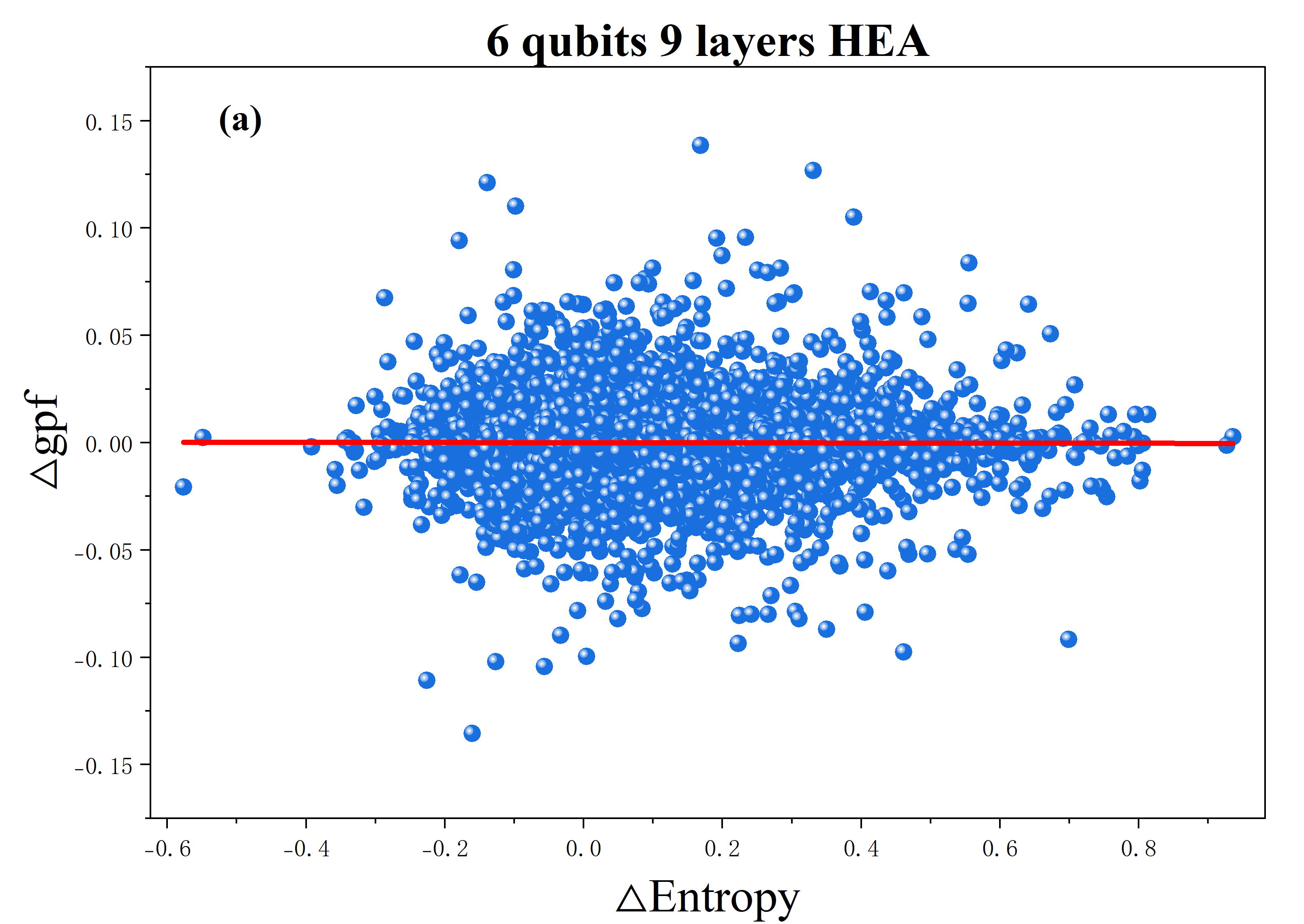}
    \includegraphics[width=0.235\textwidth]{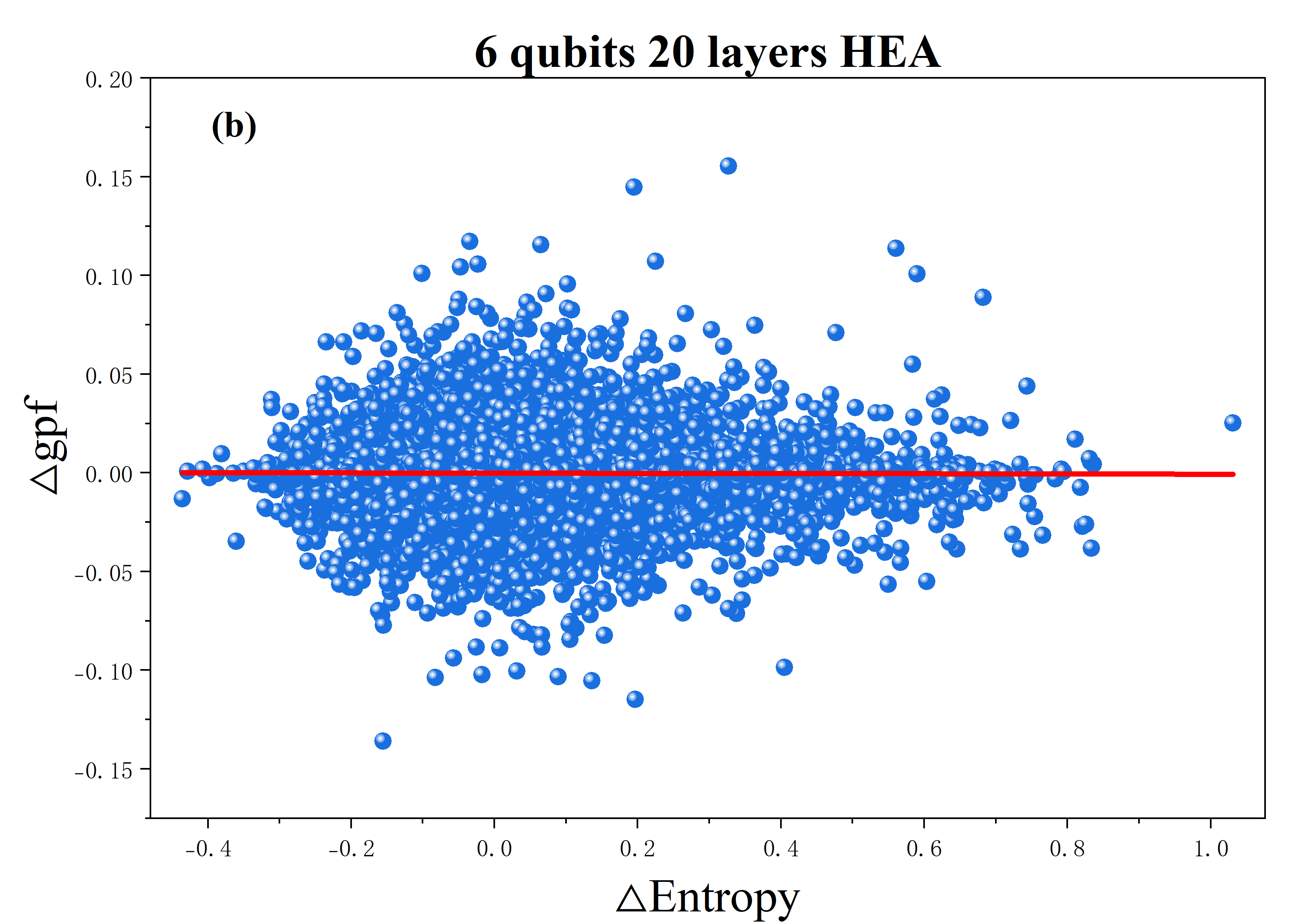}

    \vspace{0.25cm}

    \includegraphics[width=0.235\textwidth]{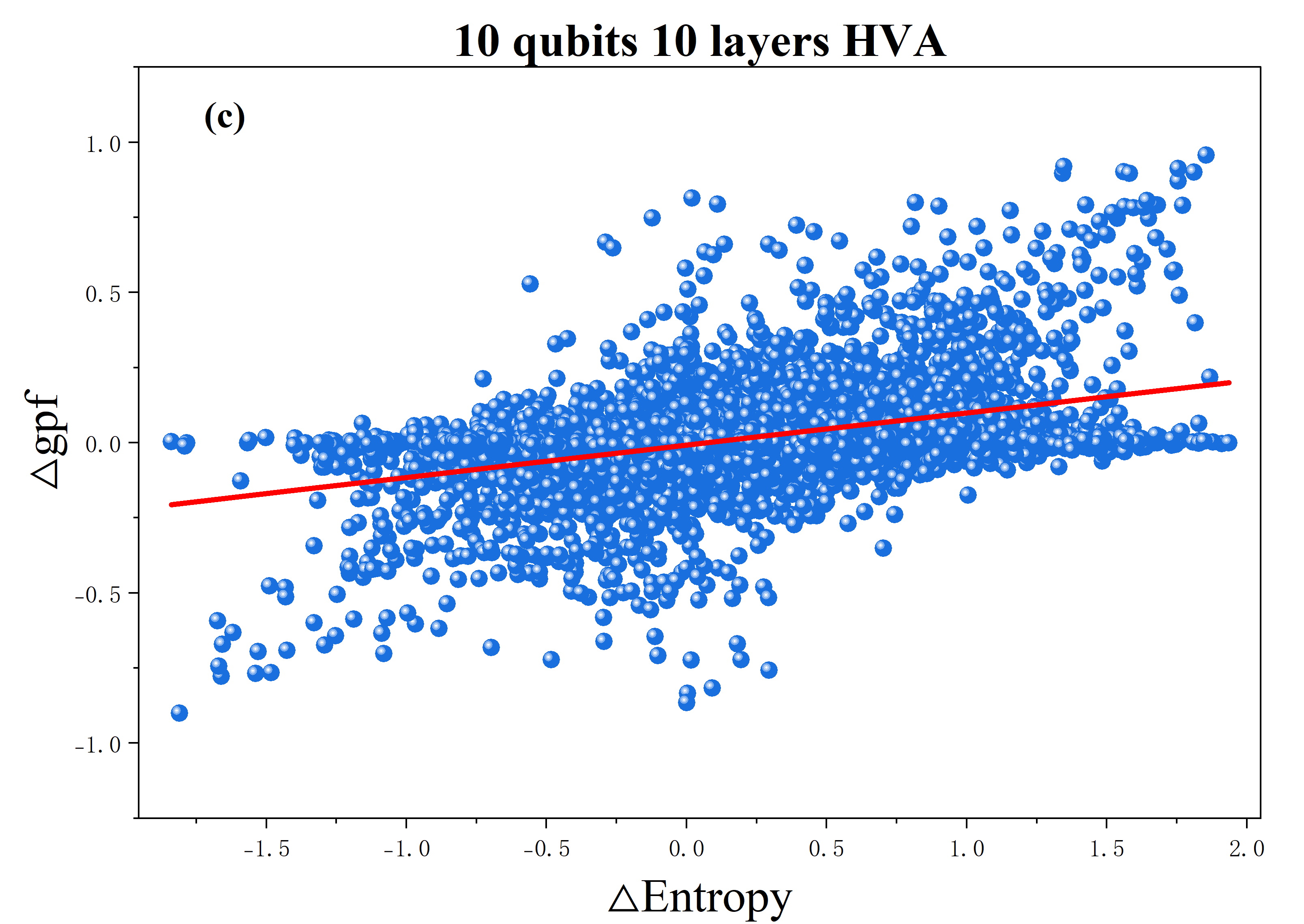}
    \includegraphics[width=0.235\textwidth]{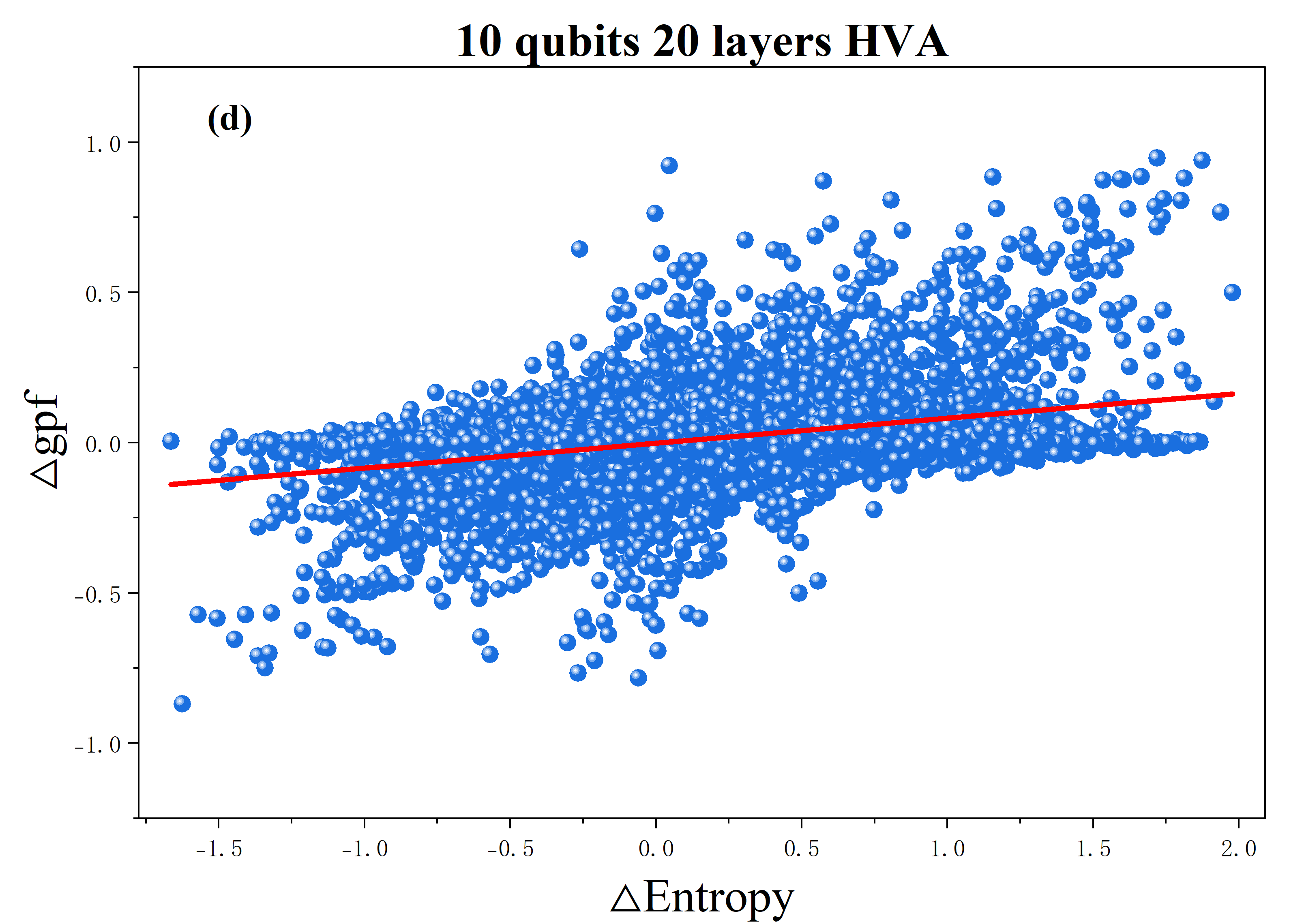}

    \caption{Statistical correlations between changes of entanglement entropy and geometric phase fraction $gpf$. The variation is calculated between adjacent layers across $10^3$ randomized HVA circuits and $10^3$ randomized HEA circuits. The red lines represent the relationship between layer-by-layer changes of the two quantities. (a) and (b) show the results of HEA circuits, while (c) and (d) show the results of HVA circuits.}
    \label{fig5}
\end{figure}

Taken together, we can infer that although the absolute magnitude of entanglement entropy influences the optimization landscape of randomized circuits, it does not exert a discernible effect on the underlying quantum state evolution. To further illustrate the specific role of entanglement in algorithm execution, we plot the step-by-step changes of entanglement entropy and geodesic distance in Fig.~\ref{fig4}, as well as the step-by-step changes of entanglement entropy and the geometric phase fraction in Fig.~\ref{fig5}. As shown in Fig.~\ref{fig4}, a clear positive correlation emerges between step-wise geodesic distance to the target space and the entanglement variations for the HVA with a structured circuit. In contrast, no such correlation is observed for the unstructured HEA circuit. This indicates that the greater the entanglement consumption, the faster the quantum states evolve, which may be masked by randomness and unstructuredness. Moreover, from Fig.~\ref{fig5}, we can also observe a positive correlation between the step-wise variation of geometric phase fraction and the entanglement entropy variations for the structured HVA circuit. A heightened rate of entanglement consumption is directly conjoined with a greater modulation of the geometric phase fraction. This verifies that in the HVA circuit, the consumption of entanglement is intrinsically coupled to the dynamical behaviors during state evolution. Conversely, the absence of such a correlation in the HEA circuit implies that its entanglement dynamics are effectively decoupled from the evolution of quantum states. Rather than reflecting a dynamical evolution, the entanglement variations in the HEA appear to originate from fluctuations in the state trajectory induced by the randomized parameterization and unstructured circuits.


\subsection{\label{sec:levelFINAL}Results of Final Optimized Circuits}

To examine whether the observed correlations persist, we further investigate the role of entanglement in optimized circuits. These circuits are obtained after performing 200 optimization iterations to achieve the lowest energy, thereby eliminating the influence of randomized parameterization.

\begin{figure}[htbp]
    \centering
    \includegraphics[width=0.35\textwidth]{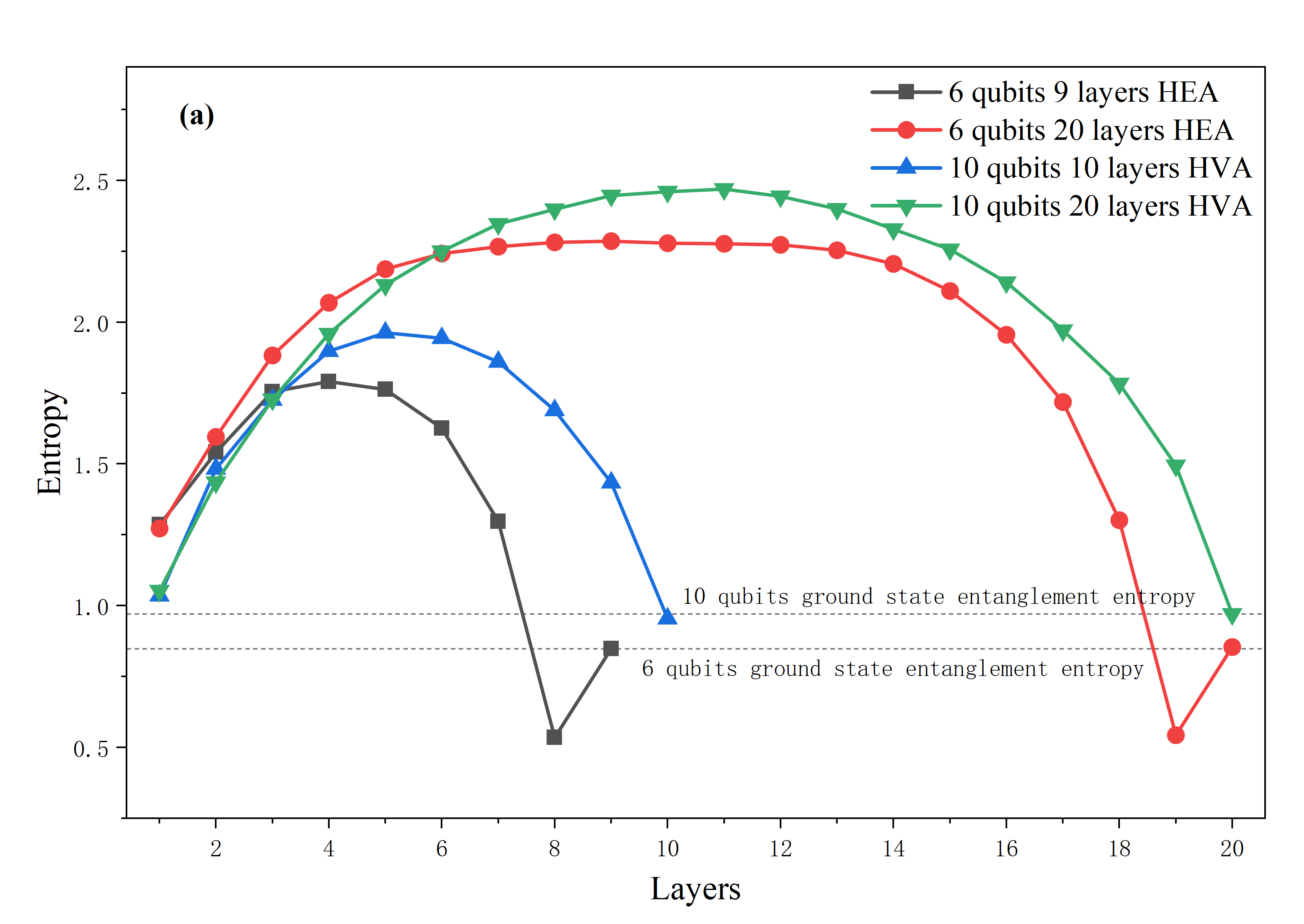}
    \includegraphics[width=0.35\textwidth]{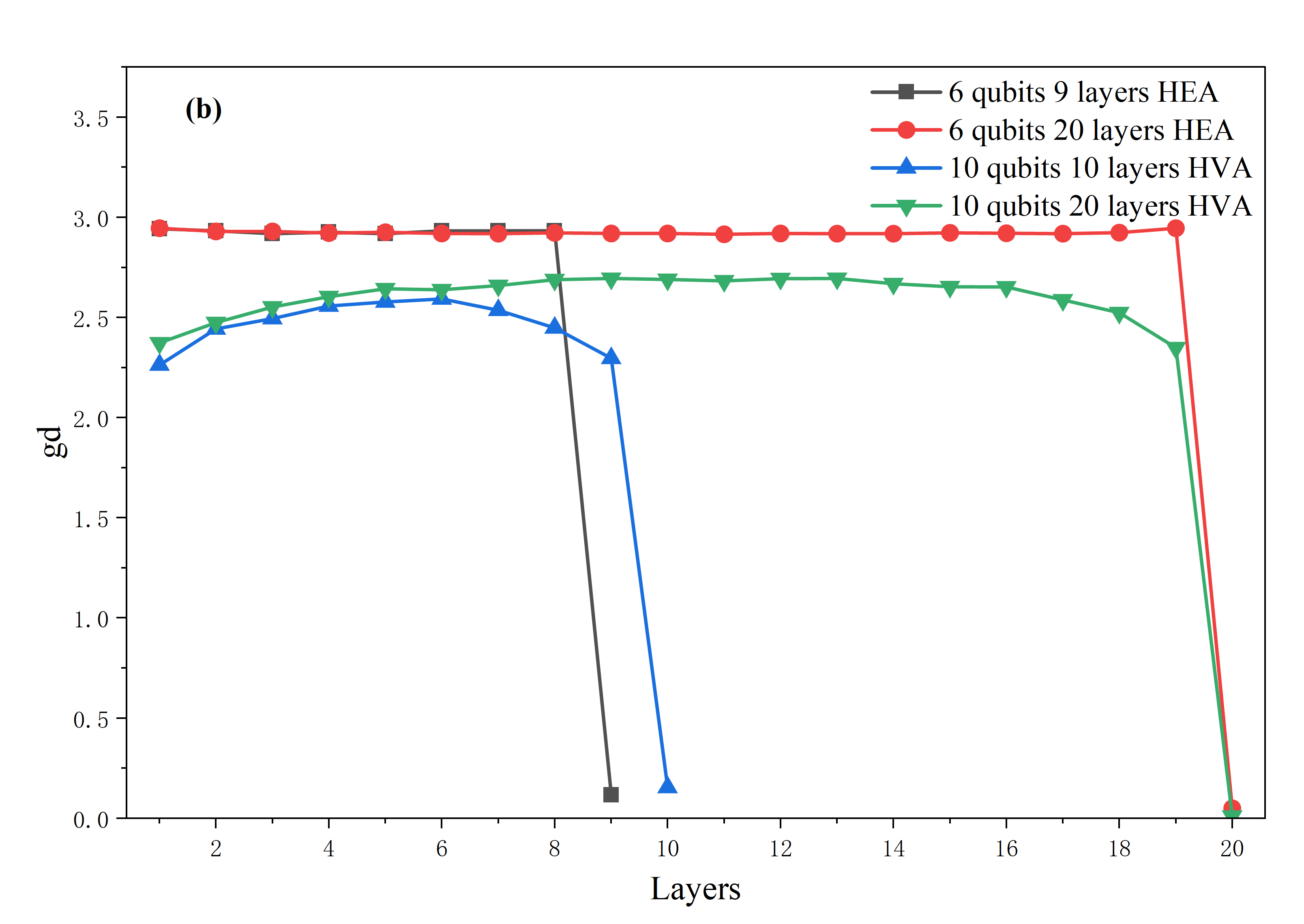}
    \includegraphics[width=0.35\textwidth]{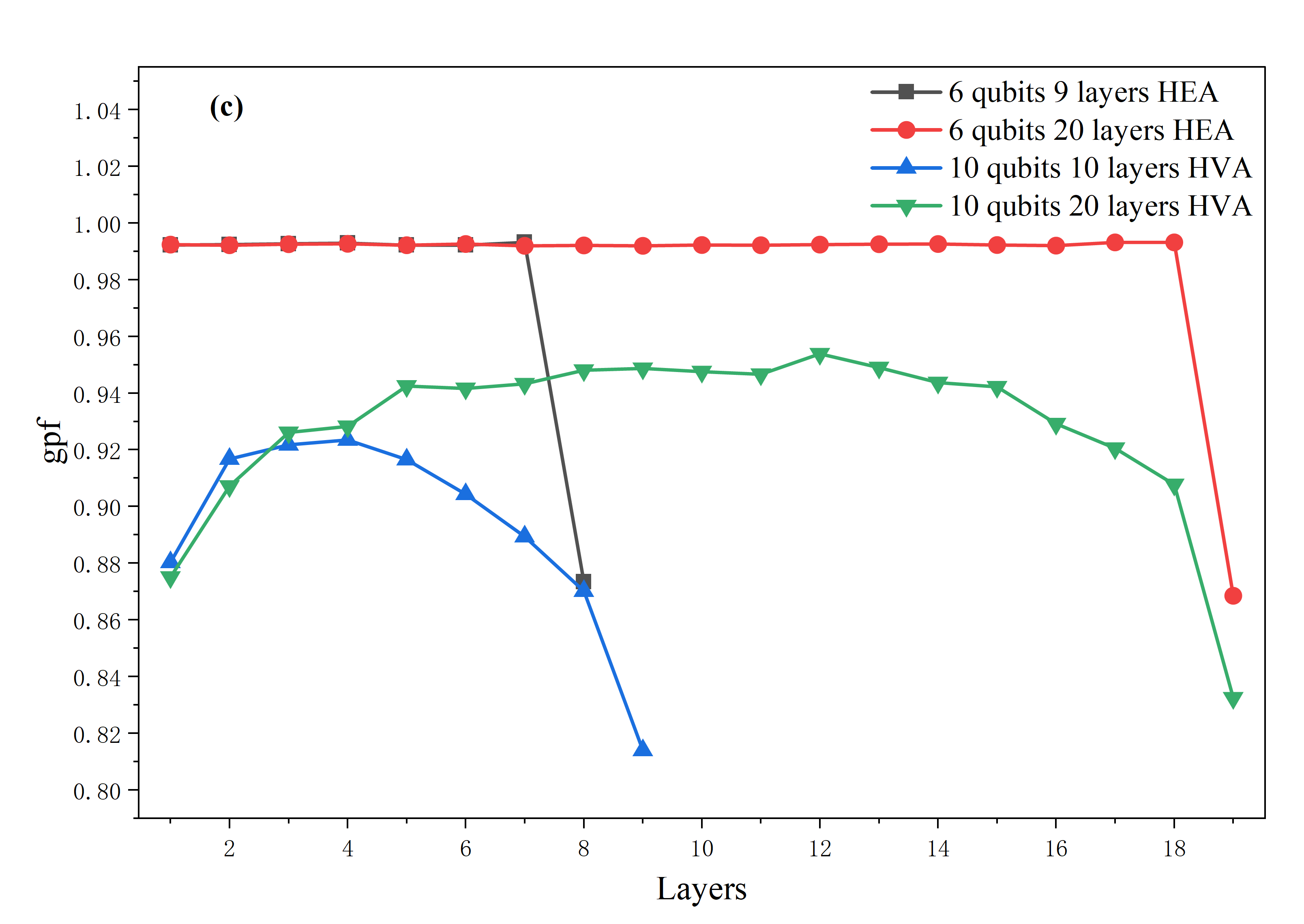}
    \caption{Data averaged $10^3$ independent trials with optimized HVA and HEA circuits. The gray line in (a) indicates the ground-state entanglement entropy calculated via exact diagonalization. (a) Variations of the bipartite entanglement entropy as a function of circuit layer. (b) Variation of geodesic distance between intermediate states and the target solution space as a function of circuit layer. (c) Variation of geometric phase fraction as a function of circuit layer.}
    \label{fig6}
\end{figure}

Fig.~\ref{fig6}(a) demonstrates that, for both HVA and HEA circuits, the entanglement entropy initially increases with circuit depth, reaches a maximum at intermediate layers, and then gradually decreases, eventually approaching the entanglement entropy of the ground state. Besides, the maximum value of entanglement entropy doesn't attain the saturation value and exhibits a slower growth. This indicates that the optimization process inherently constrains both the overall magnitude and the rate of accumulation of entanglement entropy throughout the circuit.

Surprisingly, as illustrated in Fig.~\ref{fig6}(b), the geodesic distance $gd$ towards the target space remains relatively large across almost all layers for both the HEA and HVA, except that the HVA circuit consistently exhibits a smaller geodesic distance than the HEA circuit, regardless of the number of qubits. This significant separation is maintained through the bulk of the variational layers, implying that the trajectories in parameter space remain relatively far apart until the final approach to the target region. It is only in the very last layers that the $gd$ value undergoes a sharp and substantial decline.

Fig.~\ref{fig6}(c) illustrates the geometric phase fraction evaluated at each layer. For the HEA circuit, the geometric phase remains overwhelmingly dominant throughout the entire evolution of the quantum state, with its fraction consistently approaching unity across all layers. This persistent dominance indicates that even when the influence of random parameterization is excluded, the trajectory of the HEA circuit is governed almost exclusively by the geometric configuration of the Hilbert space, with negligible contributions from dynamical factors. Such behavior is consistent with the unstructured nature of the HEA ansatz: lacking any problem-informed inductive bias that would align the evolution with the target Hamiltonian, the circuit meanders through the state manifold along paths dictated solely by the intrinsic geometry of the Hilbert space. In contrast, the HVA circuit exhibits a markedly different behavior. In the absence of randomized parameterization, the geometric phase fraction for the HVA initially increases during the early layers, yet it is subsequently suppressed and begins to decline in the intermediate layers. Although a decrease in the geometric phase fraction necessarily implies a corresponding increase in the dynamical phase contribution, it is crucial to emphasize that the geometric phase remains the dominant component throughout the entire evolution of the quantum state. With the guidance of the dynamical factor, the evolution of the quantum states in the HVA proceeds more gradually, and the final transition is not as abrupt as the HEA. Notably, the presence of the redundant layers maintains the dynamics within a regime of near-maximal geometric phase fraction, prolonging the stage during which the state evolution benefits from structured dynamical input.



\begin{figure}[htbp]
    \centering
    \includegraphics[width=0.235\textwidth]{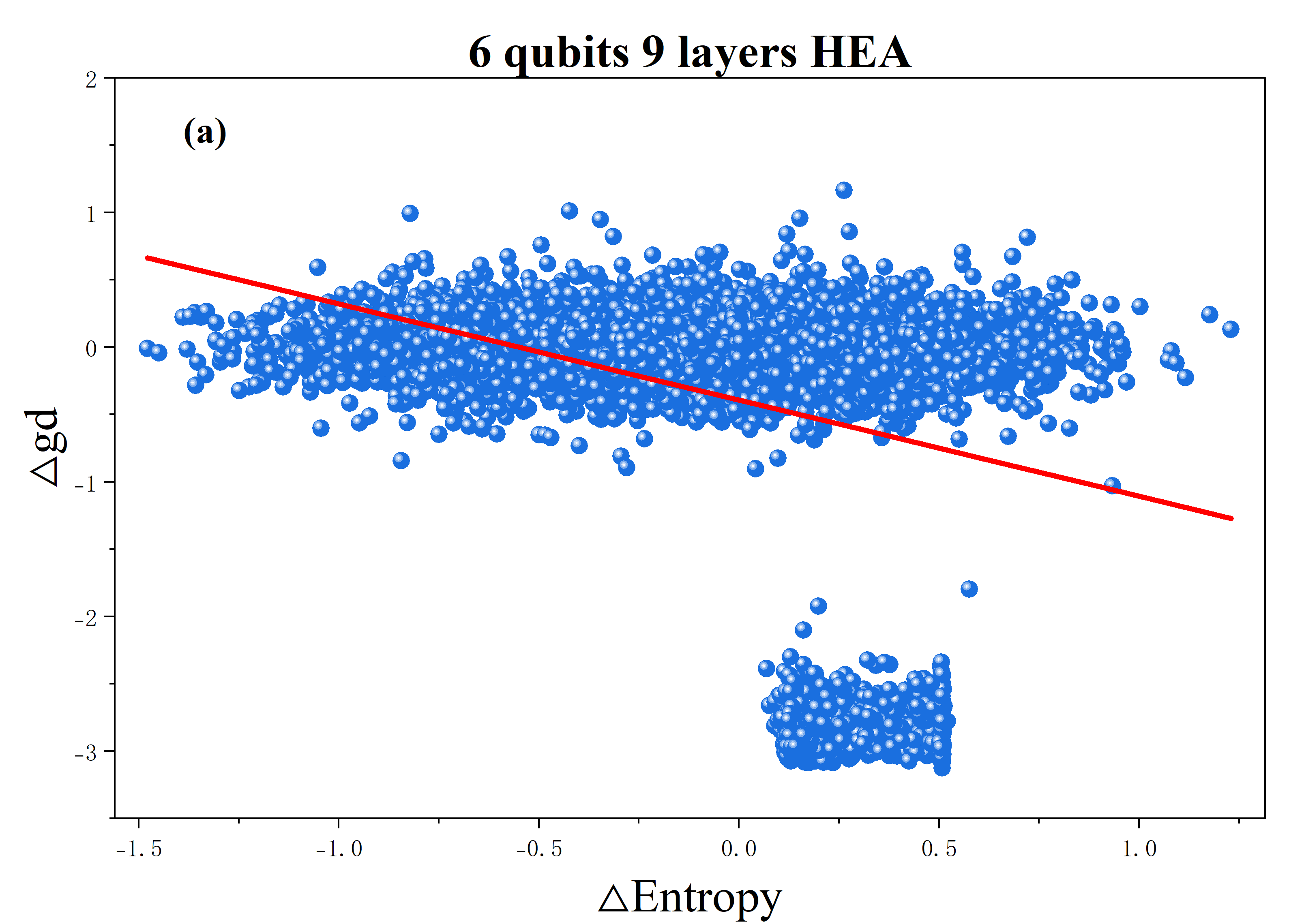}
    \includegraphics[width=0.235\textwidth]{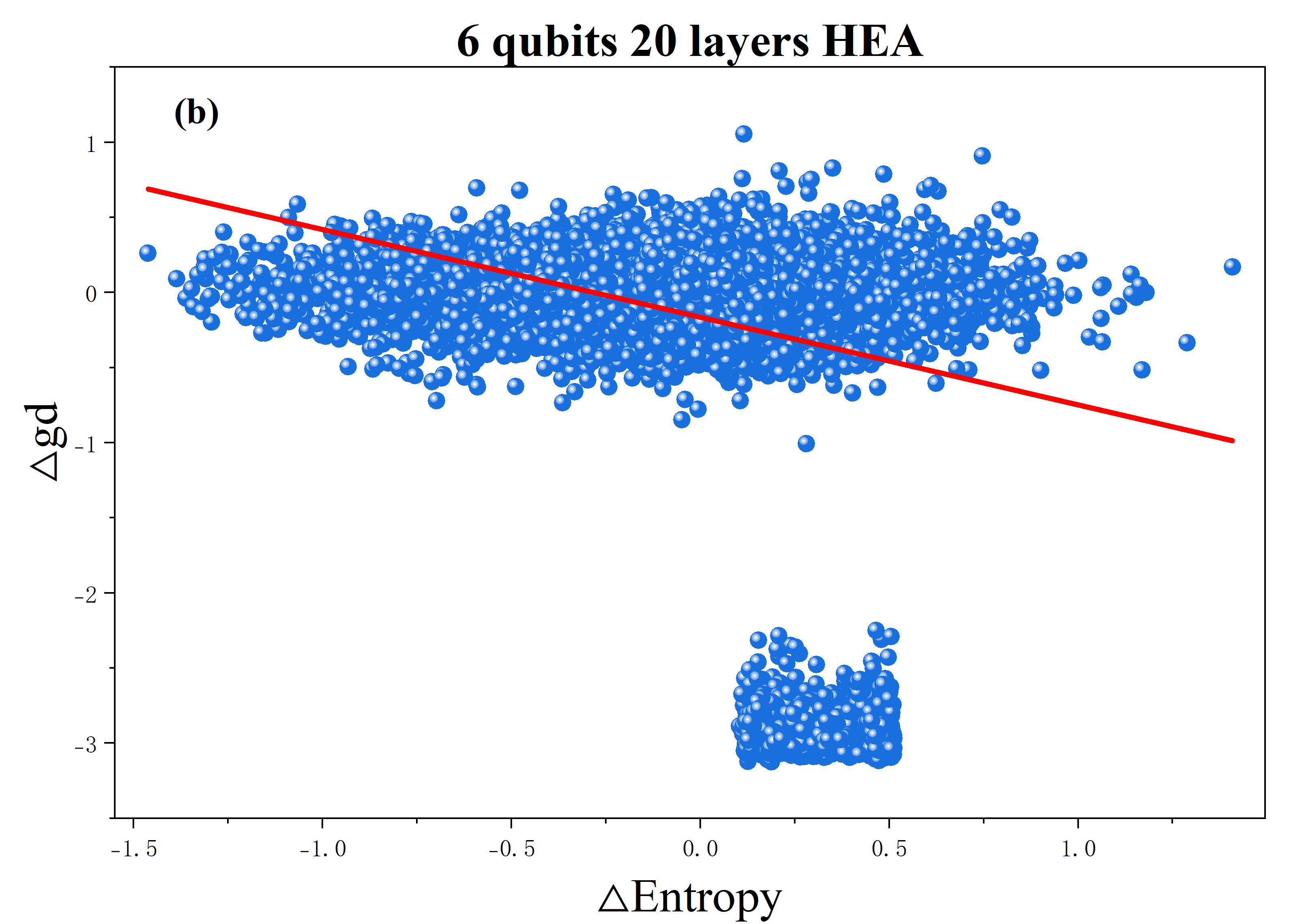}

    \vspace{0.25cm}

    \includegraphics[width=0.235\textwidth]{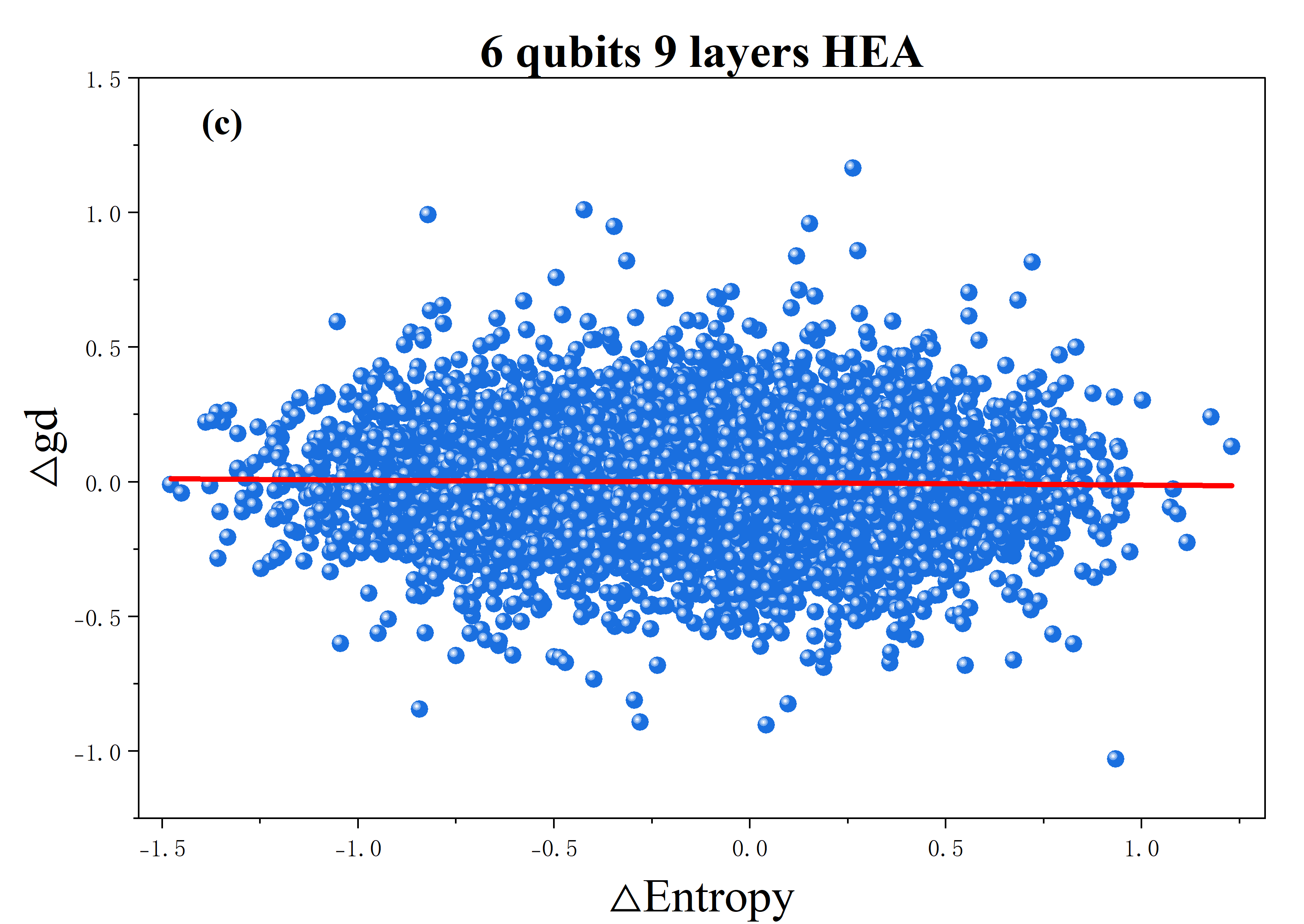}
    \includegraphics[width=0.235\textwidth]{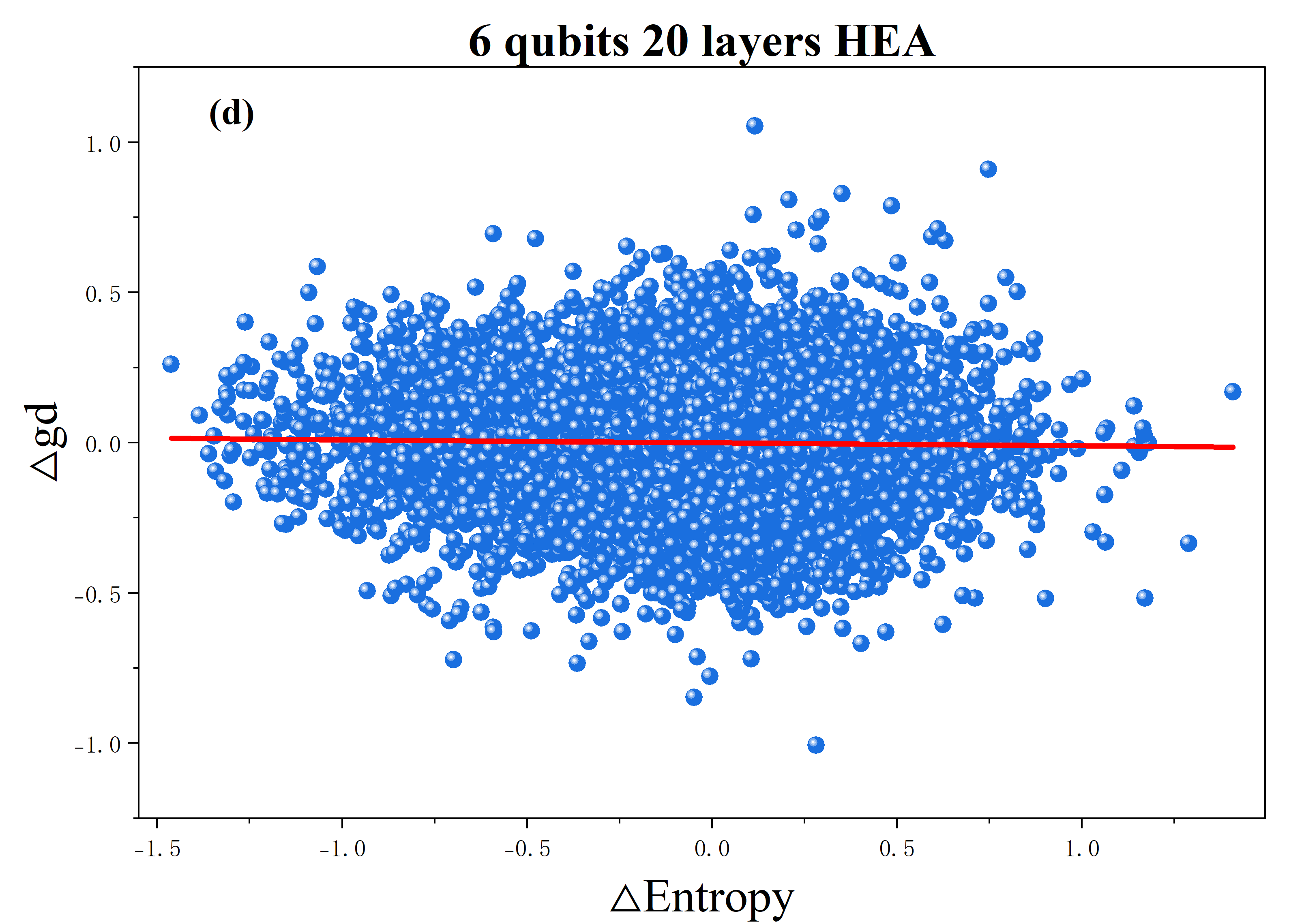}

    \vspace{0.25cm}

    \includegraphics[width=0.235\textwidth]{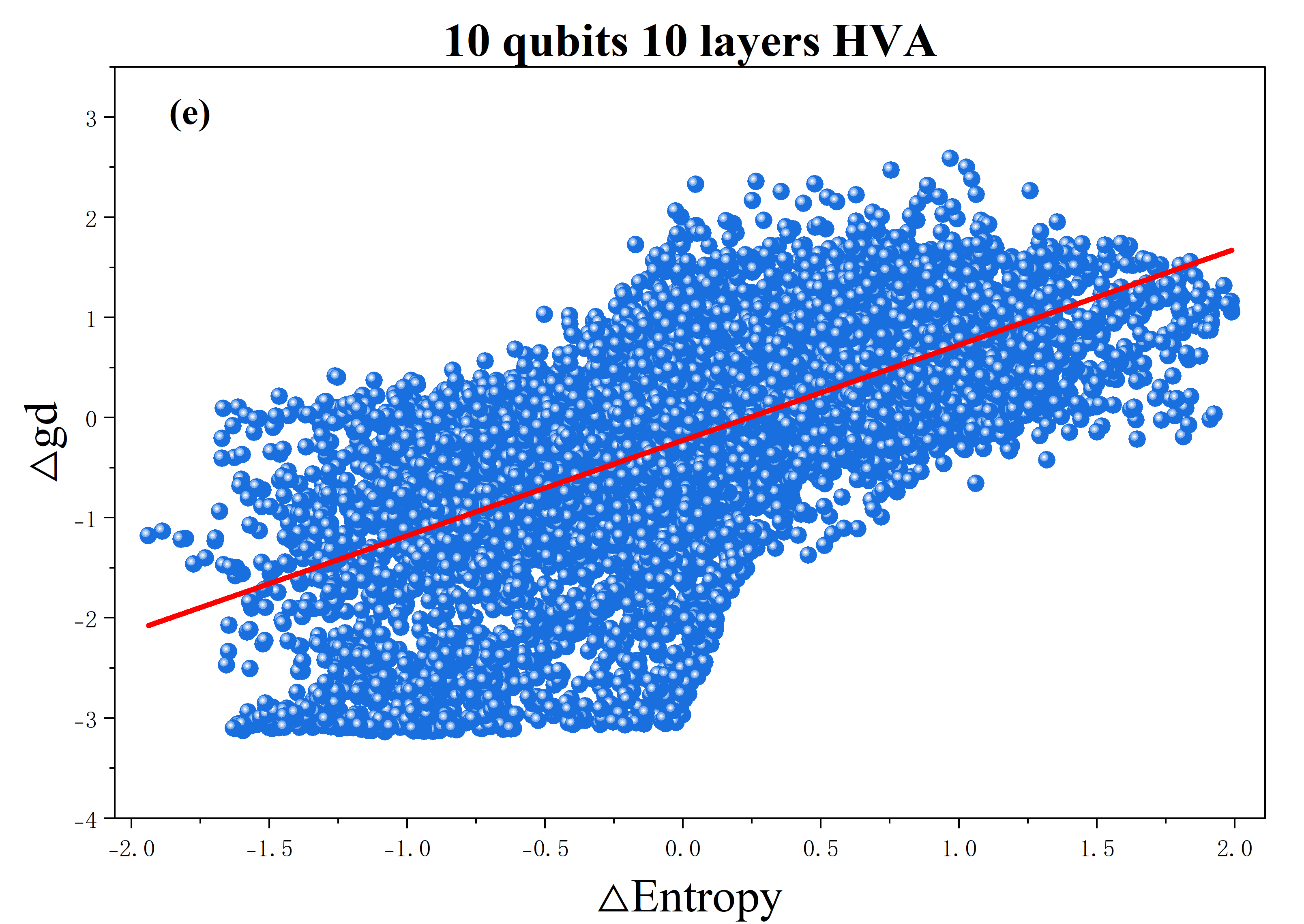}
    \includegraphics[width=0.235\textwidth]{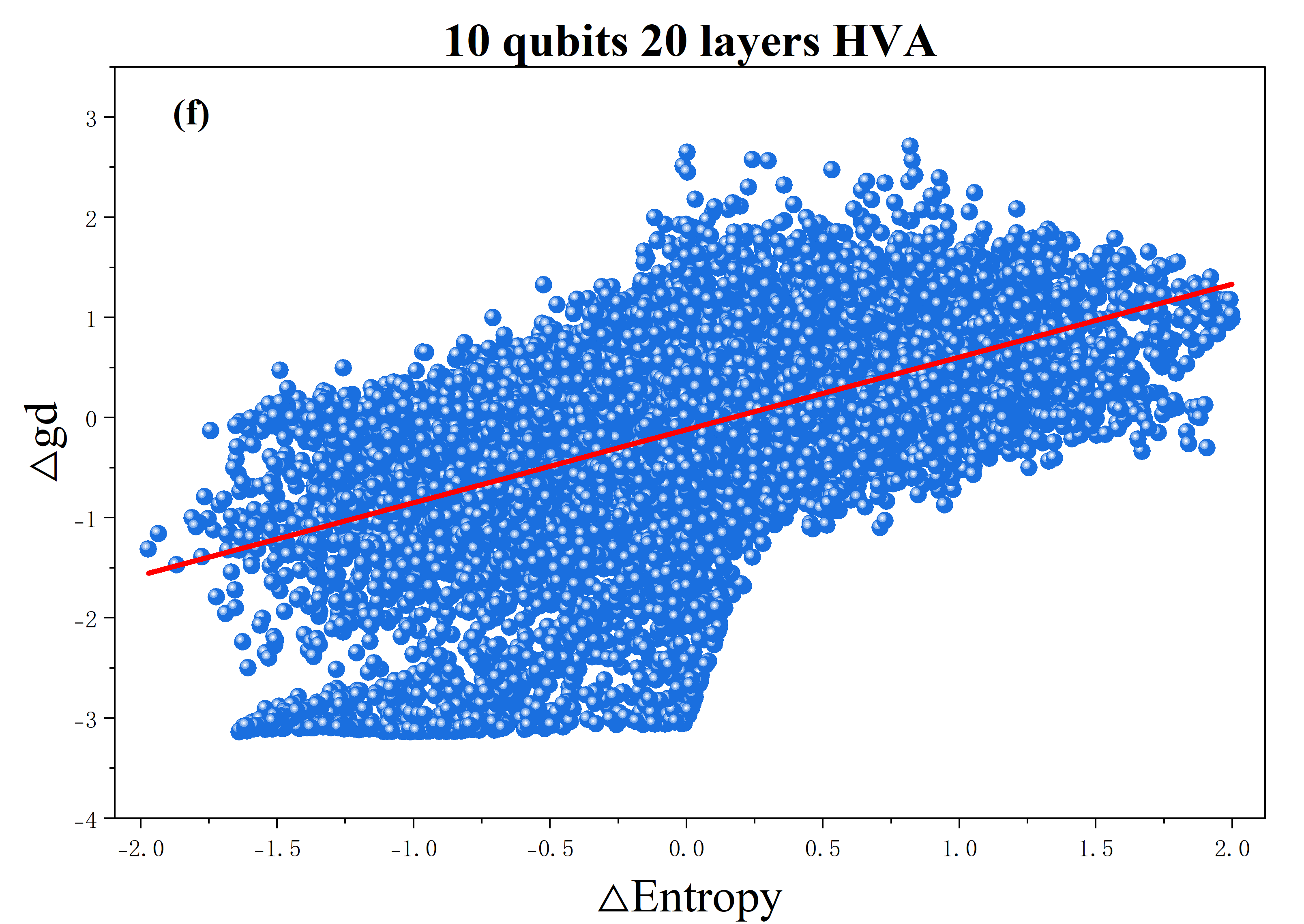}

    \caption{Statistical correlations between changes of entanglement entropy and geodesic distance $gd$ toward the target space. The variation is calculated between adjacent layers across $10^3$ optimized HVA circuits and $10^3$ optimized HEA circuits. The red lines represent the relationship between layer-by-layer changes of the two quantities. (a) and (b) show the results of HEA circuits, while (c) and (d) show the same results but with the final layer’s data removed. (e) and (f) show the results of HVA circuits.}
    \label{fig7}
\end{figure}

The relationship between changes in entanglement entropy and geodesic distance is explored in Fig.~\ref{fig7}. After isolating the effects of randomized parameterization, HVA exhibits a more pronounced positive correlation between step-wise geodesic distance to the target space and the entanglement variations than that observed in randomized circuits. This implies a more efficient utilization of entanglement. Rather than dissipating entanglement through stochastic fluctuations in the state trajectory, the structured ansatz of the HVA channels the consumption of entanglement into coherent, goal-directed evolution toward the target manifold.  Notably, in optimized HEA circuits, a negative correlation between the step-wise variation of geodesic distance and the entanglement entropy variations is observed. Combined with Fig.~\ref{fig6}, we find that this stems primarily from the behavior of the final layer, which is characterized by a very small increment in entanglement and a substantial jump in geodesic distance. This singular feature exerts a disproportionate influence on the overall correlation analysis, artificially skewing the trend toward a negative value. It is therefore essential to exclude the data from the final layer when assessing the intrinsic relationship between entanglement dynamics and state evolution. Indeed, once the final-layer data are removed from the analysis, no meaningful correlation persists, as demonstrated in Fig.~\ref{fig7}(c)(d). This absence of correlation mirrors the behavior observed in randomly initialized HEA circuits. Collectively, these observations indicate that even after optimization, HEA circuits remain fundamentally incapable of harnessing entanglement as a functional resource to drive the evolution of the quantum state.



\begin{figure}[htbp]
    \centering
    \includegraphics[width=0.235\textwidth]{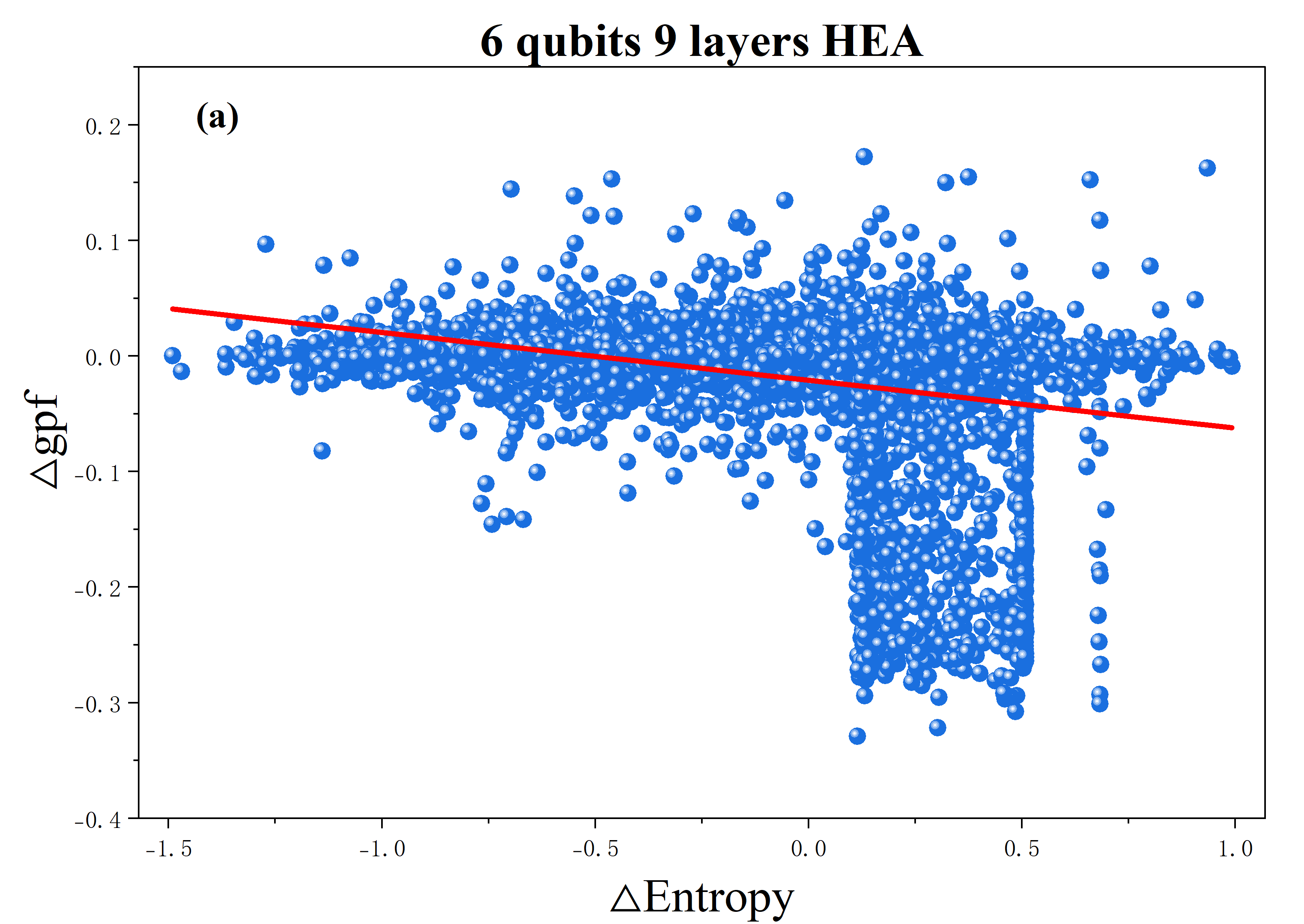}
    \includegraphics[width=0.235\textwidth]{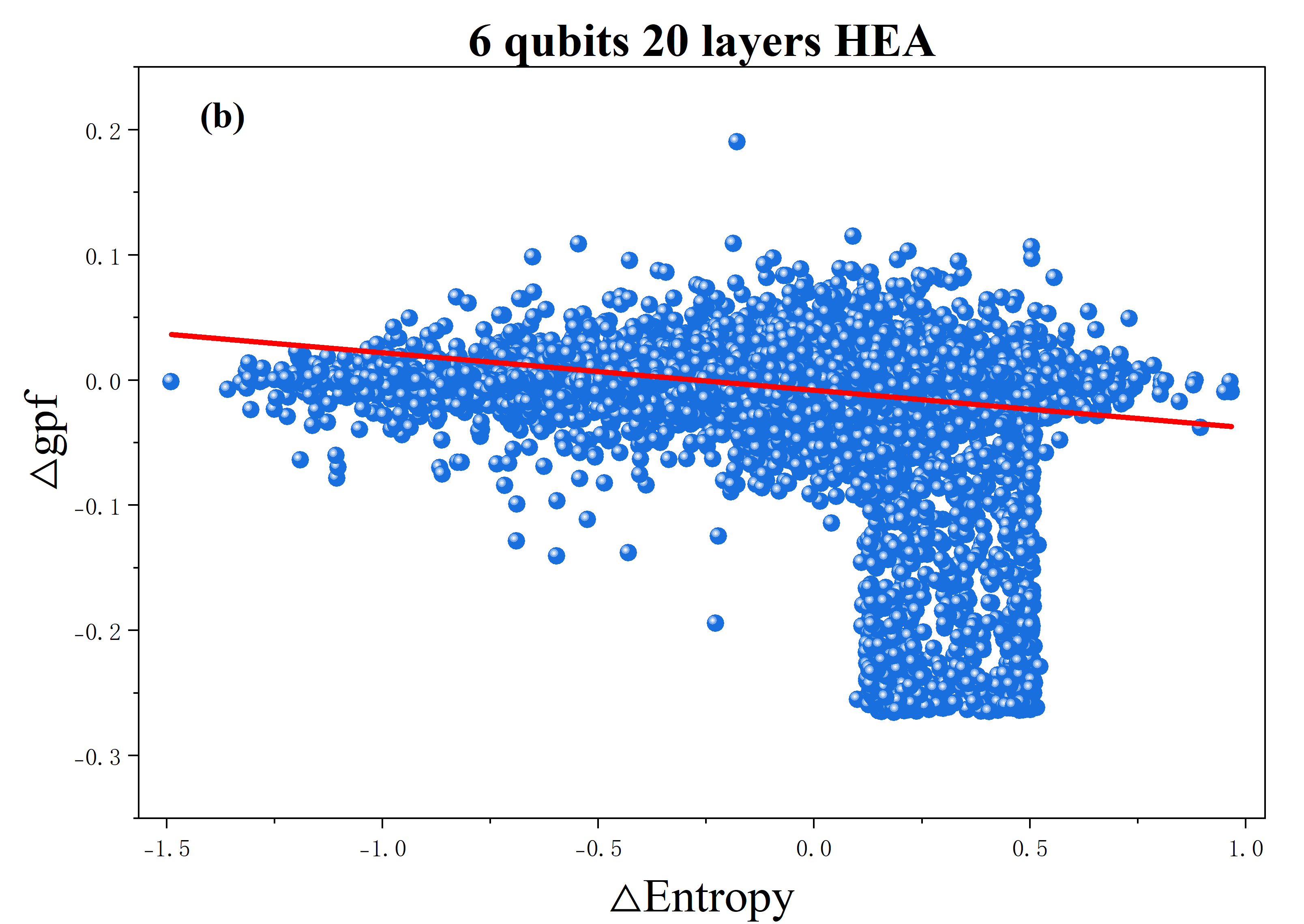}

    \vspace{0.25cm}

    \includegraphics[width=0.235\textwidth]{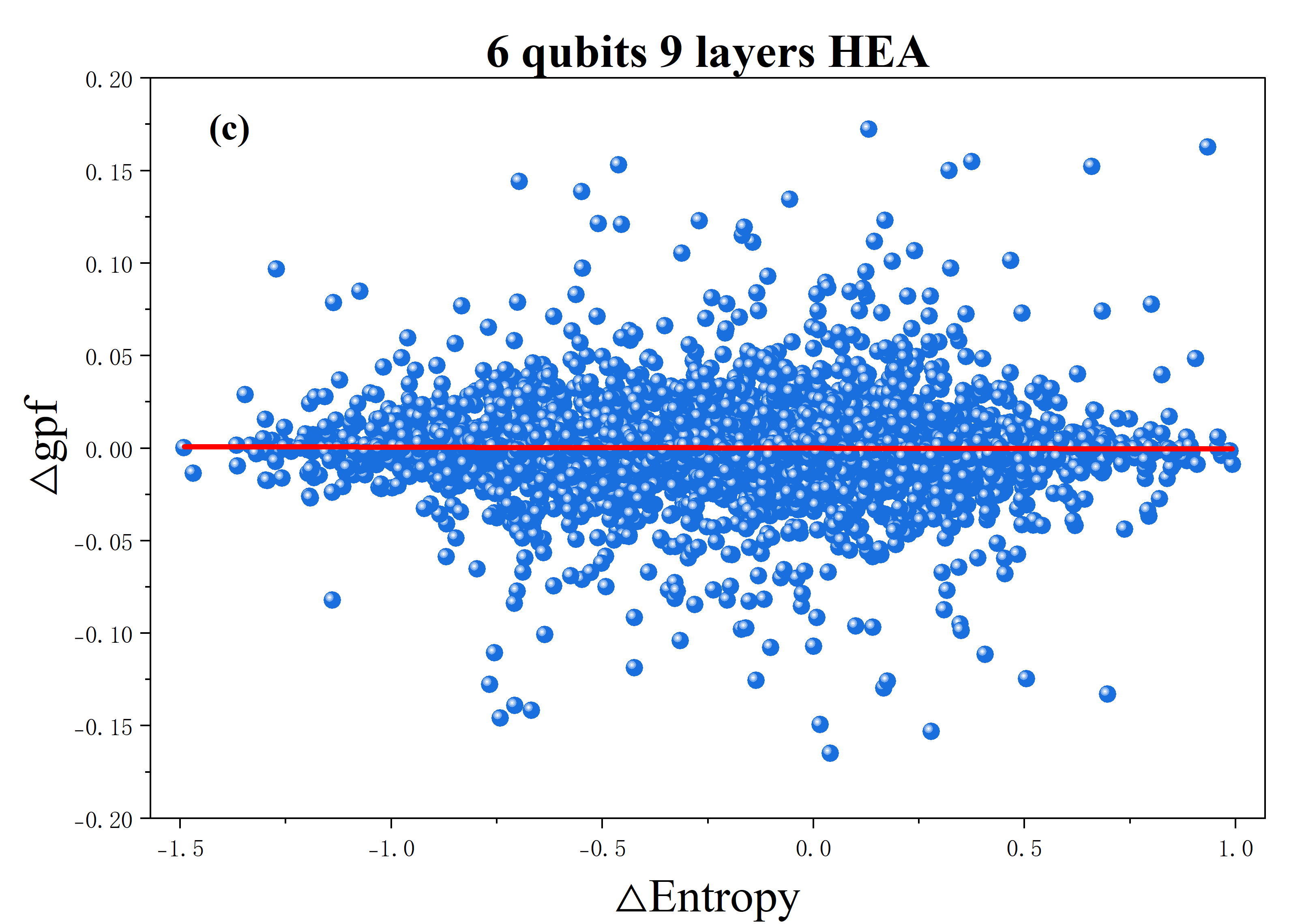}
    \includegraphics[width=0.235\textwidth]{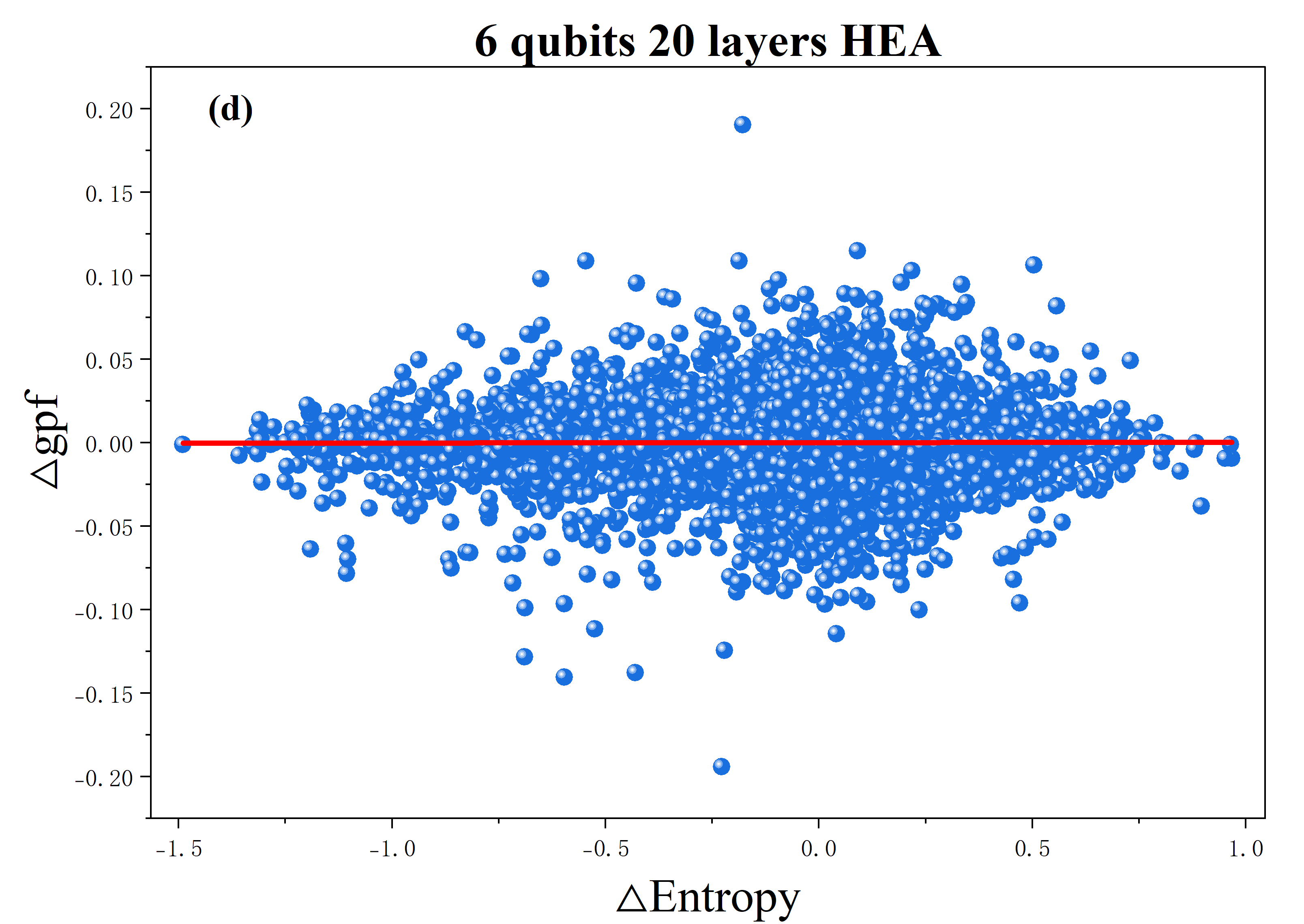}

    \vspace{0.25cm}

    \includegraphics[width=0.235\textwidth]{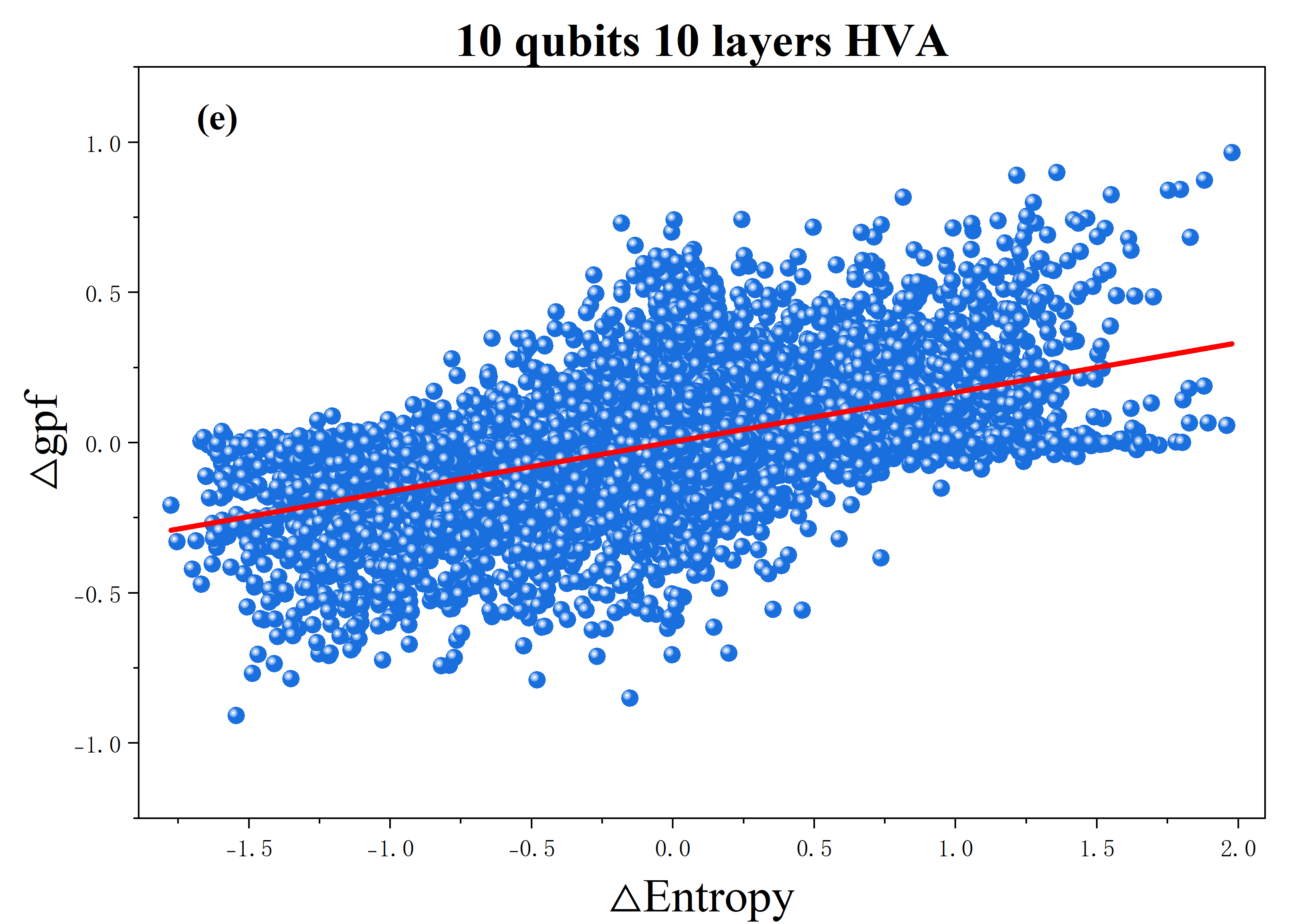}
    \includegraphics[width=0.235\textwidth]{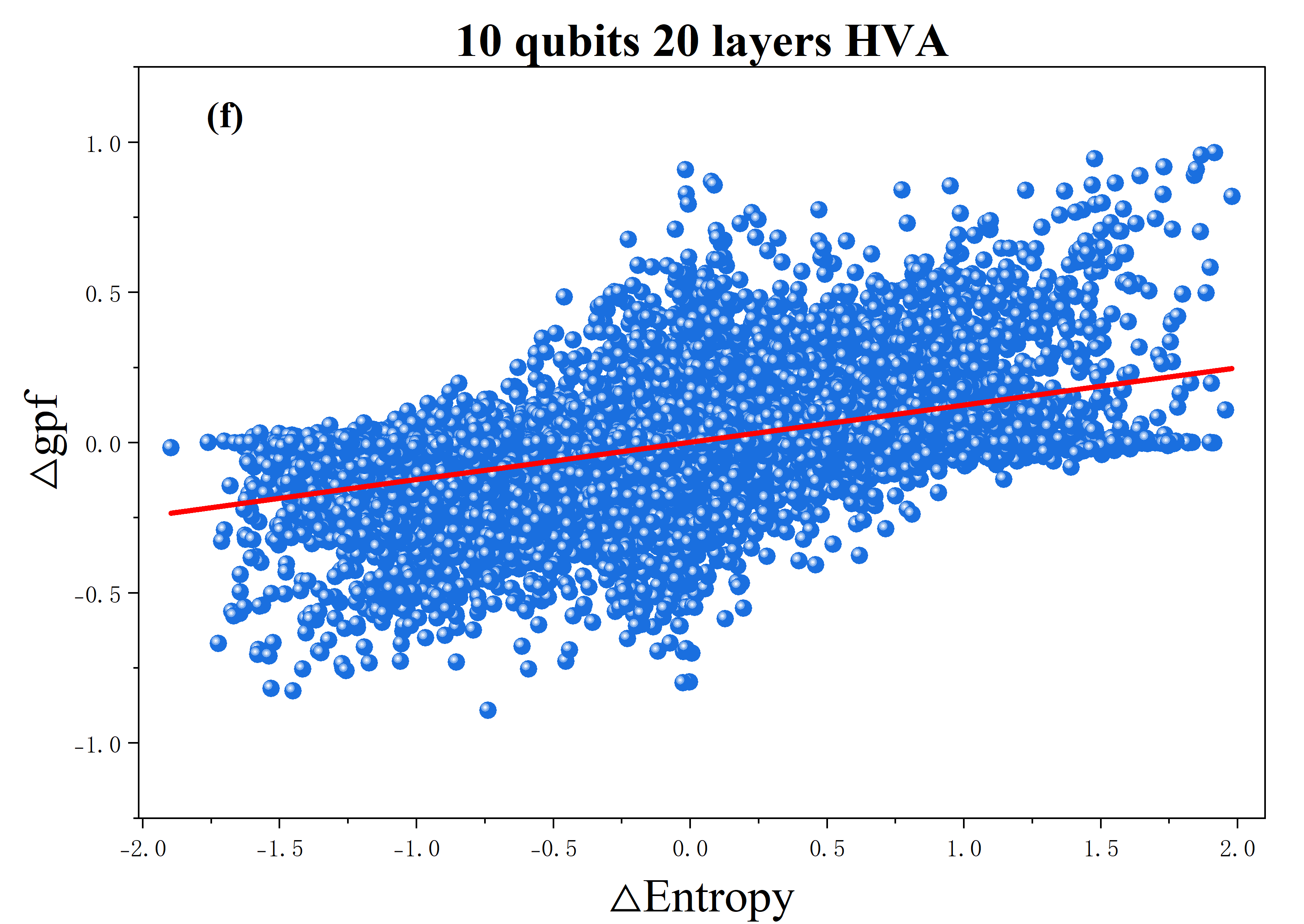}

    \caption{Statistical correlations between changes of entanglement entropy and geometric phase fraction $gpf$. The variation is calculated between adjacent layers across $10^3$ optimized HVA circuits and $10^3$ optimized HEA circuits. The red lines represent the relationship between layer-by-layer changes of the two quantities. (a) and (b) show the results of HEA circuits, while (c) and (d) show the same results but with the final layer’s data removed. (e) and (f) show the results of HVA circuits.}
    \label{fig8}
\end{figure}

Fig.~\ref{fig8} explores the relationship between the step-by-step changes of entanglement entropy and the geometric phase fraction. It is clear that after optimization, the HVA also exhibits a more pronounced positive correlation between these two quantities when the randomness in the circuit parameters is removed. This persistent correspondence reinforces the conclusion that, within the HVA framework, the consumption of entanglement is intrinsically coupled to the dynamical behaviors during state evolution. Specifically, the problem-inspired architecture of the HVA endows the circuit with an inherent capacity to harness entanglement as a functional resource, translating its consumption into measurable dynamical effects. Conversely, in the HEA case, even after careful parameter optimization, any meaningful correlation between entanglement entropy dynamics and phase variation remains absent (reasons as stated in the previous paragraph). The fact that this coupling fails to emerge even after parameter optimization indicates that it is not merely a fortuitous byproduct of random initialization, but rather a fundamental structural property of the HEA ansatz itself. The problem-agnostic nature of the HEA architecture inherently decouples entanglement variations from the evolution of the quantum states.

\section{\label{sec:levelC}CONCLUSION}

In this work, we have investigated the role of entanglement in variational quantum algorithm executions from a geometric perspective, contrasting the dynamical behavior of problem-inspired HVA architecture and problem-agnostic HEA architecture. By tracing the layer-wise evolution of entanglement entropy, the corresponding geodesic distances towards the target space, and the step-by-step geometric phase fraction, we have uncovered a fundamental distinction in how these ansatz architectures utilize quantum entanglement.

Our analysis reveals that in HEA circuits, the evolution of the quantum state is overwhelmingly dominated by the geometric phase, with the trajectory dictated almost exclusively by the curvature of the Hilbert space. In this regime, entanglement dynamics and state evolution are effectively decoupled: variations in entanglement entropy exhibit no meaningful correlation with the changes in geodesic distance and geometric phase fraction, regardless of whether the parameters are randomly initialized or fully optimized. This decoupling indicates that, despite generating substantial entanglement during algorithm execution, the unstructured nature of HEA precludes the utilization of entanglement as a functional dynamical resource. 

In contrast, HVA circuits exhibit an enhanced contribution from the dynamical phase, which aligns the state evolution with the structure of the target Hamiltonian. Consequently, a robust positive correlation emerges between the consumption of entanglement and the rate of quantum state evolution. In the HVA framework, greater consumption of entanglement directly translates into accelerated variation of geodesic distance and geometric phase fraction. This intrinsic coupling persists under both randomized and optimized parameterization, underscoring that the problem-informed inductive bias of HVA endows the circuit with an inherent capacity to harness entanglement as a coherent, goal-directed resource.

Our geometric framework provides a novel diagnostic tool for understanding and potentially guiding the design of variational quantum circuits. It is suggested that ansatz architectures should aim not merely to generate entanglement, but to guide its consumption in a manner that actively drives the state toward the target manifold.

\begin{acknowledgments}
This work is supported by the National Natural Science Foundation of China under Grants No.61975005, Beijing Academy of Quantum Information Science under Grants No.Y18G28, and the Fundamental Research Funds for the Central Universities.
\end{acknowledgments}

\nocite{*}

\bibliography{apssamp}

\end{document}